\documentclass[twocolumn, nofootinbib, superscriptaddress, amsmath,amssymb,aps,prx,nobalancelastpage]{revtex4-2}

\usepackage{amssymb}
\usepackage{amsfonts}
\usepackage{comment}
\usepackage{graphicx}
\usepackage{dcolumn}
\usepackage{bm}
\usepackage{amsmath}
\usepackage{upgreek}
\usepackage[colorlinks,linkcolor=red,citecolor=blue,urlcolor=red]{hyperref}
\usepackage{geometry}
\usepackage{physics}
\usepackage{bbm}
\usepackage{xpatch}
\usepackage{orcidlink}

\newcommand\mbf[1]{\mathbf{#1}}
\newcommand{\argmax}{\mathop{\mathrm{argmax}}\limits}
\newcommand{\argmin}{\mathop{\mathrm{argmin}}\limits}

\newcommand{\V}{\mathbb{V}}

\begin{document}

\title{Towards Quantum-Limited Spatial Resolution of NV-Diamond Magnetometry}

\author{Nico Deshler\,\orcidlink{0000-0001-8657-3237}}
\email{ndeshler@arizona.edu}
\affiliation{Wyant College of Optical Sciences, University of Arizona, Tucson, AZ, USA}

\author{Declan Daly}
\affiliation{Electrical and Computer Engineering, University of Maryland, College Park, MD, USA}

\author{Ayan Majumder}
\affiliation{Department of Electrical Engineering, Indian Institute of Technology Bombay, Mumbai, India}

\author{Kasturi Saha}
\affiliation{Department of Electrical Engineering, Indian Institute of Technology Bombay, Mumbai, India}

\author{Saikat Guha}
\email{saikat@umd.edu}
\affiliation{Electrical and Computer Engineering, University of Maryland, College Park, MD, USA}
\affiliation{Wyant College of Optical Sciences, University of Arizona, Tucson, AZ, USA}

\begin{abstract}

    Optically addressable ensembles of solid-state defects, such as nitrogen vacancy (NV) centers, are a leading modality for imaging-based magnetometry, thermometry and strain sensing. However, monitoring the fluorescence of individual defects within a sub-diffraction ensemble remains an outstanding challenge that currently limits access to atomic-scale features and dynamics. For compact clusters of NVs, we formulate imaging-based atomic sensing as a low-dimensional multiparameter estimation task in which one seeks to localize each defect and quantify the field strength in its immediate vicinity. In this work, we employ optical spatial mode demultiplexing (SPADE) to enhance localization and brightness estimation accuracy at sub-diffraction scales. Specifically, we develop a two-stage sensing protocol that augments direct imaging by projecting the incoming optical field onto point spread function (PSF)-adapted, i.e., PAD spatial modes and Yuen-Kennedy-Lax (YKL) spatial modes enabling efficient extraction of emitter positions and brightnesses. The YKL-SPADE measurement employed for brightness estimation is shown to be quantum-optimal in the case of two emitters and establishes a new connection between quantum detection and estimation theories. We numerically evaluate the statistical performance of our protocol for sub-diffraction optically detected magnetic resonance (ODMR) and Rabi sensing experiments. Compared to conventional focal plane intensity measurements, our protocol improves emitter localization accuracy by 6$\times$ and brightness estimation accuracy by 2$\times$ for tightly confined ensembles, residing well below the diffraction limit.
\end{abstract}

\maketitle

\section*{Introduction}
\label{sec: Introduction}
Solid state quantum sensors, such as nitrogen vacancy (NV) and silicon vacancy (Si-V) centers in diamond, exploit the electronic energy level structure of lattice defects to measure a variety of physical quantities on a microscopic scale, including magnetic fields, electric fields, strain, and temperature \cite{levine_principles_2019, barry_sensitivity_2020, schirhagl_nitrogen-vacancy_2014, rajendran_method_2017}. By transducing a physical quantity of interest into the optical domain, these devices have enabled precision metrology across various domains including electrical engineering \cite{reed_machine_2024, lenz_hardware_2024, oliver_vector_2022}, biology \cite{le_sage_optical_2013, zhang_toward_2021, balasubramanian_nitrogen-vacancy_2014}, chemical physics \cite{devience_nanoscale_2015, mamin_nanoscale_2013, glenn_high-resolution_2018}, condensed matter \cite{ku_imaging_2020}, and astrophysics \cite{marshall2021directional}. However, diamond samples may contain individual defects that are spaced by mere nanometers. Resolving features at these scales with conventional imaging techniques becomes prohibitively difficult due to diffraction \cite{levine_principles_2019,scholten_widefield_2021}, which currently limits the spatial resolution of high-performance confocal microscopes to approximately $200$ nanometers \cite{arai_fourier_2015,guo_wide-field_2024}.  Thus, super-resolution imaging methods are needed for nanoscale quantum sensing applications, particularly in biology and chemistry \cite{abobeih_atomic-scale_2019, lovchinsky_nuclear_2016}, as well as quantum simulation and control \cite{cai_large-scale_2013}. 

Super-resolution imaging aspires to faithfully discern features in a scene that are smaller than the Rayleigh limit $\sigma \sim f \lambda/D$, which is characterized by the effective focal length $f$, entrance pupil diameter $D$, and operating wavelength $\lambda$ of the imaging system. Currently, state-of-the-art sensing techniques that circumvent the diffraction limit require actively modulating the ambient environment of the sample and/or the probe illumination. For example, in Fourier magnetic imaging \cite{arai_fourier_2015, guo_wide-field_2024} a known magnetic field gradient is introduced over the sample that stratifies the precession rate of different defects based on their position. Using timed $\pi$-pulse sequences, the position of individual defects can be inferred from the photoluminescent response of each defect as a function of pulse timing. Alternative methods such as spin-RESOLFT sensing (REversible Saturable OpticaL Fluorescence Transitions) illuminate the solid state material with a structured doughnut beam that suppresses emission from neighboring defects around a sub-diffraction target \cite{jaskula_superresolution_2017, boretti_nitrogen-vacancy_2019, chen_subdiffraction_2015}. Although these methods have enabled extreme resolutions down to the $10$'s of nanometers \cite{Amawi:2024_3DFourierMagneticImaging,jaskula_superresolution_2017}, we are unaware of any super-resolution sensing techniques that invoke passive measurement strategies to remain largely non-invasive to the sample environment. Passive sensing methods may be particularly useful for sensing living biological samples with photothermal and magnetic fragility \cite{Zablotskii:2016_Cell-Life_MagneticFields}. 

\begin{figure*}
  \centering
    \includegraphics[width=\linewidth]{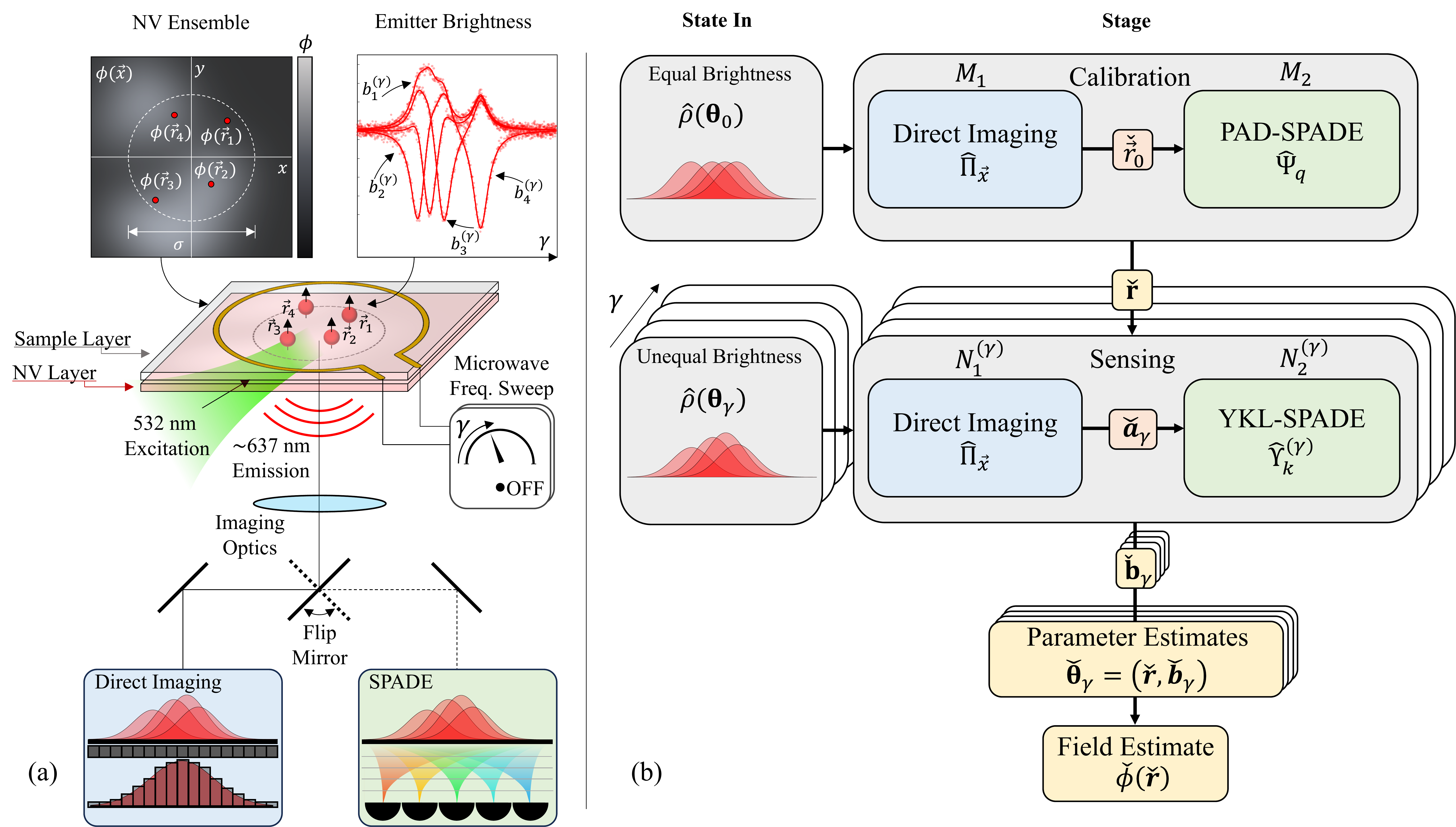}
    \caption{ \textbf{(a)} Schematic of our imaging-based NV sensing apparatus configured for an ODMR experiment. The goal of the apparatus is to estimate the spatially-varying magnetic field $\phi(\vec{x})$ in the sample layer at the location of the NVs $\vec{r}_k$. As shown, all NVs reside within a diffraction-limited spot diameter $\sigma$ such that their pair-wise separations are sub-diffraction. Each emitter has a relative brightness $b^{(\gamma)}_k$ which varies as a function of microwave drive frequency $\gamma$. The setup consists of two distinct measurement modules - a direct imaging focal plane detector array and a reconfigurable SPADE device. The beam path can be directed to either module through an actuated flip mirror. \textbf{(b)} Diagram of the proposed two-stage protocol for achieving passive super-resolution sensing with solid-state defects. During the calibration stage, all emitters are driven to fluoresce with equal brightness such that the only unknown parameters in the received state $\hat{\rho}(\bm{\theta}_{0})$ are the emitter locations. An estimate of the emitter locations $\check{\mbf{r}}$ is made from a combination of direct imaging and PAD-SPADE measurements. These location estimates are then fed to the sensing stage wherein an estimate of the brightnesses $\check{\bm{b}}_{\gamma}$ is made from a combination of direct imaging and YKL-SPADE measurements under the modulated state $\hat{\rho}(\bm{\theta}_{\gamma})$. The collection of brightness estimates made for each microwave frequency $\gamma$ is then used to reconstruct the field evaluated at the emitter locations $\check{\phi}(\check{\mbf{r}})$.}
    \label{fig: Sensing Protocol Schematic}
\end{figure*}

Since Tsang \textit{et al.} \cite{Tsang:2016} demonstrated the optimality of spatial mode demultiplexing (SPADE) for attaining the quantum Crame\'r-Rao bound on precision estimation of distance between two incoherent point sources, a wellspring of theoretical and experimental publications have emerged that apply SPADE to various optical sensing tasks including object discrimination and change detection \cite{Grace:2022_SubDiffraction_Object_Discrimination,Grace:2023_QUSUM,Lu:2018_QuantumDetection_1v2_Sources}, astronomical imaging \cite{Deshler:2025_QuantumOptimalCoronagraph,Deshler:2025_ExoplanetDiscoveryQuantumLimits,HuangLupo:2021_HypothesisTestingExoplanets,Sajjad:2024_LongBaselinQFI,Padilla:2025_LongbaselineDistributedEntanglement,KitLee:2023}, precision spectroscopy \cite{Donohue:2018_TimeFrequencySPADE,Huang:2023_ExoplanetSpectroscopy}, optomechanical sensing \cite{Pluchar:2025_ImagingQuantumOptomechanics,Tsang:2023_QuantumNoiseSpectroscopy,Choi:2024_ResonatorImaging}, displacement sensing \cite{Boucher:2020_DisplacementSensing,He:2024_BeamPositionSensing,Sajjad:2021_ObjectLocalization},
 and extended-object imaging \cite{Zanforlin:2022_SuperresolutionImagingHypothesisTesting, Rehacek:2017_RealisticSuperresolution, Dutton:2019_ObjectLength, Duplinskiy:2025_StructuredIllumination_SPADE, Prasad:2019_3DSourcePair,Tsang:2019_QuantumBoundsIncoherentImaging,Matlin:2022_ExtendedObjectSPADE,Bearne:2021_ConfocalSPADE}. The efficacy of SPADE lies in its ability to extract task-specific information about a scene encoded in the full transverse phase-amplitude profile of the optical field arriving at the image plane. Depending on the task at hand, particular spatial mode bases may exhibit enhanced sensitivity to parameters of interest carried by the optical field. This opens an opportunity for measurement optimization \textit{viz.} mode basis selection.
 
 In this paper, we consider the novel sensing architecture shown in Fig. \ref{fig: Sensing Protocol Schematic}(a) which achieves passive super-resolution sensing with atomic vacancies in diamond by invoking spatial mode sorting. Specifically, we propose a simple two-stage protocol that consists of a calibration stage to localize the vacancies followed by a sensing stage to estimate their brightnesses. Each stage consists of a direct imaging measurement and a SPADE measurement shown in Fig. \ref{fig: Sensing Protocol Schematic}(b). In comparison to conventional focal plane intensity measurements, our protocol enhances localization accuracy of sub-diffraction ensembles by $6\times$ and brightness estimation accuracy by $2\times$ using equivalent photon resources. We further show that these improvements in localization and brightness estimation accuracy translate to dramatic improvements in the accuracy of sub-diffraction field reconstructions under both DC and AC sensing paradigms.
 
\section*{Protocol for Super-resolution Vacancy Sensing}
\label{sec: Protocol}

\subsection*{Sensing Model}
\label{sec: Sensing Model}
Imaging-based quantum sensing endeavors to resolve some spatially-varying scalar field $\phi(\vec{x}_{\,})$ (e.g. magnetic, temperature, strain, etc.) where $\vec{x} = (x,y)$ is a 2D coordinate in the plane of the vacancy sample. The field is determined indirectly through its influence on the electronic energy level structure of the vacancies. Changes in the energy levels of a vacancy manifest as a photoluminescent (intensity) response $I(\gamma|\phi)$ with respect to experimentally controlled modulation parameter $\gamma$ (e.g. microwave frequency, detuning, pulse spacing, pulse number, pulse duration, etc). For instance, in optically detected magnetic resonance (ODMR), the local magnetic field around an NV center introduces Zeeman splitting of degenerate electronic spin states. The splitting energy depends linearly on the local magnetic field strength and can be measured by sweeping the frequency of a microwave probe while monitoring the NV intensity. While the independent variable of a given experiment might vary to suit the target application, the vast majority of color vacancy experiments feature intensity or contrast as a dependent variable. In this regard, the accuracy of the field estimate is contingent on the accuracy of the intensity estimates.

Here we consider an ensemble of $K$ emitters located at $\{\vec{r}_k \}_{k=1}^{K}$ with intensities $I_k^{(\gamma)} = I(\gamma|\phi(\vec{r}_k))$. We define $\vec{r}_0 = \frac{1}{K}\sum_{k=1}^{K}\vec{r}_k$ to be the geometric center of the ensemble and $I_0^{(\gamma)} = \sum_{k=1}^{K}I_k^{(\gamma)}$ to be the total intensity. This work is concerned with the sub-diffraction ensembles - that is, configurations where the Euclidean distance between every emitter and the geometric midpoint is less than half the Rayleigh limit $|\vec{r}_{k}-\vec{r}_0|<\sigma/2$. Our primary interest is in the \textit{relative intensity} of each emitter $b_k^{(\gamma)} \equiv I_k^{(\gamma)}/I_0^{(\gamma)}$, which we hereafter refer to as the \textit{brightness}. The complete set of parameters describing the emitter ensemble for any given $\gamma$ is thus,
\begin{subequations}
    \begin{align}
    \bm{\theta}_{\gamma} &= [\mbf{r},\mbf{b}_{\gamma}] \in \Theta \subset \mathbb{R}^{K\times 3},
    \label{eqn: Estimation Parameters}\\
    \mbf{r} &\equiv [\vec{r}_1,\ldots,\vec{r}_K]^{\intercal} \in \mathbb{R}^{K\times 2}, \,{\text{and}}\\
    \mbf{b}_{\gamma} &\equiv [b_1^{(\gamma)},\ldots,b_K^{(\gamma)}]^{\intercal} \in  \mathcal{S}_{K} \subset \mathbb{R}^{K},
    \end{align}
\end{subequations}
where $\mathcal{S}_{K} = \{\mbf{b}\in \mathbb{R}^{K}: \sum_{k=1}^{K}b_k = 1, b_k>0\}$ denotes the standard probability simplex with the boundary omitted (i.e. all emitters are assumed to have non-zero brightness). Formally, the full parameter space is thus an open set $\Theta =  \mathbb{R}^{K\times 2} \times \mathcal{S}_K$. Approximating the emitters as weak incoherent quasi-monochromatic thermal sources \cite{Tsang:2016}, the signal-bearing single-photon state of the optical field arriving at the image plane can be written as a parametric density operator,
\begin{equation}
\begin{split}
\rho(\bm{\theta}_{\gamma}) &= \rho(\mbf{r},\mbf{b}_{\gamma}) = \sum_{k=1}^{K}b_k^{(\gamma)} \dyad{\psi(\vec{r}_k)},
\end{split}
\label{eqn: Density Operator}
\end{equation}
where $\ket{\psi(\vec{r}_k)} \equiv \int_{\mathbb{R}^2}d\vec{x} \, \psi(\vec{x}-\vec{r}_k)\ket{\vec{x}}$ is a single-photon state of the square-normalized complex point spread function (PSF) $\psi(\vec{x})$ of the imaging system  centered at the emitter location $\vec{r}_k$. Additionally, $\ket{\vec{x}} = \hat{a}^{\dagger}(\vec{x})\ket{0}$ is a single-photon state at the position $\vec{x}$ with $\braket{\vec{x}}{\vec{x}'} = \delta(\vec{x}-\vec{x}')$, and $d\vec{x} \equiv dxdy$ denotes the 2D differential area element.\footnote{Throughout this work, we restrict our attention to the multimode Hilbert space $\mathcal{H}$ consisting of single-photon states of all transverse scalar electric field modes at the image plane. This space is isomorphic to the space of square-normalizeable complex functions in two dimensions $\mathbb{L}^2(\mathbb{R}^2,\mathbb{C})$ equipped with the standard inner product $\braket{u}{v} = \int_{\mathbb{R}^2} d\vec{x}\, u^*(\vec{x})v(\vec{x})$ for all $\ket{u},\ket{v}\in\mathcal{H}$. In the main text, we reserve $\hat{I}$ for the identity operator on $\mathcal{H}$.}

The fundamental barrier to brightness estimation at sub-diffraction scales arises from the overlap between states of the optical field received from different emitters $\braket{\psi(\vec{r}_j)}{\psi(\vec{r}_{k})}\neq \delta_{jk}$. Consequently, any physically-realizable single-copy measurement has a non-zero probability of erroneously assigning a detected photon to a particular emitter of origin \cite{Nielsen_Chuang:2010}. Thus, our ability to estimate the brightness $b_k$ of the $k^{\rm th}$ emitter is inexorably connected to our ability to discriminate the state $\ket{\psi_k}$ from the other emitter states. 

To expound this intuition, consider the opposing case where the states $\ket{\psi(\vec{r}_k)}$ are orthogonal, which is approximately true when the emitters are spatially resolved (i.e., when $|\vec{r}_j-\vec{r}_k| \gg \sigma$ for all $j\neq k$). In this case, the detection of a photon in the $k^{\rm th}$ state $\ket{\psi(\vec{r}_k)}$ definitively indicates that the photon originated from the $k^{\rm th}$ emitter. Therefore, an optimal estimate of each brightness parameter $b_k$ may be constructed by measuring multiple copies of the single-photon state $\rho(\bm{\theta})$ in the orthonormal basis defined by $\{\ket{\psi(\vec{r}_k)}\}$ and reporting the fraction of photons observed in each state. 

When the states are not orthogonal, such as when the emitters are unresolved, a detected photon may not be definitively associated with one particular emitter.\footnote{This problem persists even if the emitter positions $\vec{r}_k$ are known exactly, which is already a formidable hurdle for sub-diffraction sensing.} Thus, we expect the optimal measurement for discriminating the states $\{\ket{\psi(\vec{r}_k)}\}$ to also be an informative single-copy measurement for estimation of the brightness parameters. This perspective allows us to leverage tools from quantum hypothesis testing to construct spatial mode projections that constitute optimal state discrimination measurements. However, unlike the standard problem formulation for quantum hypothesis testing, here the prior probabilities $b_k$ are unknown, which demands formulating a pre-estimate of the brightnesses in order to construct the optimal measurement.

Our protocol consists of two sequential stages shown in Fig. \ref{fig: Sensing Protocol Schematic}(b).  The `calibration stage' determines the locations $\vec{r}_k$ of the emitters present in the field of view.\footnote{The total number of emitters $K$ is assumed to be known \textit{a priori}. Generalizing to an unknown number of emitters requires treating $K$ as an unknown parameter and running multiple competing estimation models in parallel. We point the curious reader to \cite{KitLee:2023} which demonstrates the simultaneous estimation of the number of emitters and their locations under the constraint that $1\leq K\leq K_{\text{max}}$ for some user-defined $K_{\text{max}}$.} Throughout this stage we force the emitters to radiate with equal brightness by optically pumping the vacancy sample in the absence of the field $\phi$. The `sensing stage' estimates the relative brightness of each emitter $b_k^{(\gamma)}$ for every $\gamma$. Critically, the brightness estimates are \textit{conditioned on our knowledge of the positions of the emitters acquired in the calibration stage}. In this regard, our approach is one of sequential estimation.

\begin{figure*}
 \centering
\includegraphics[width=\linewidth]{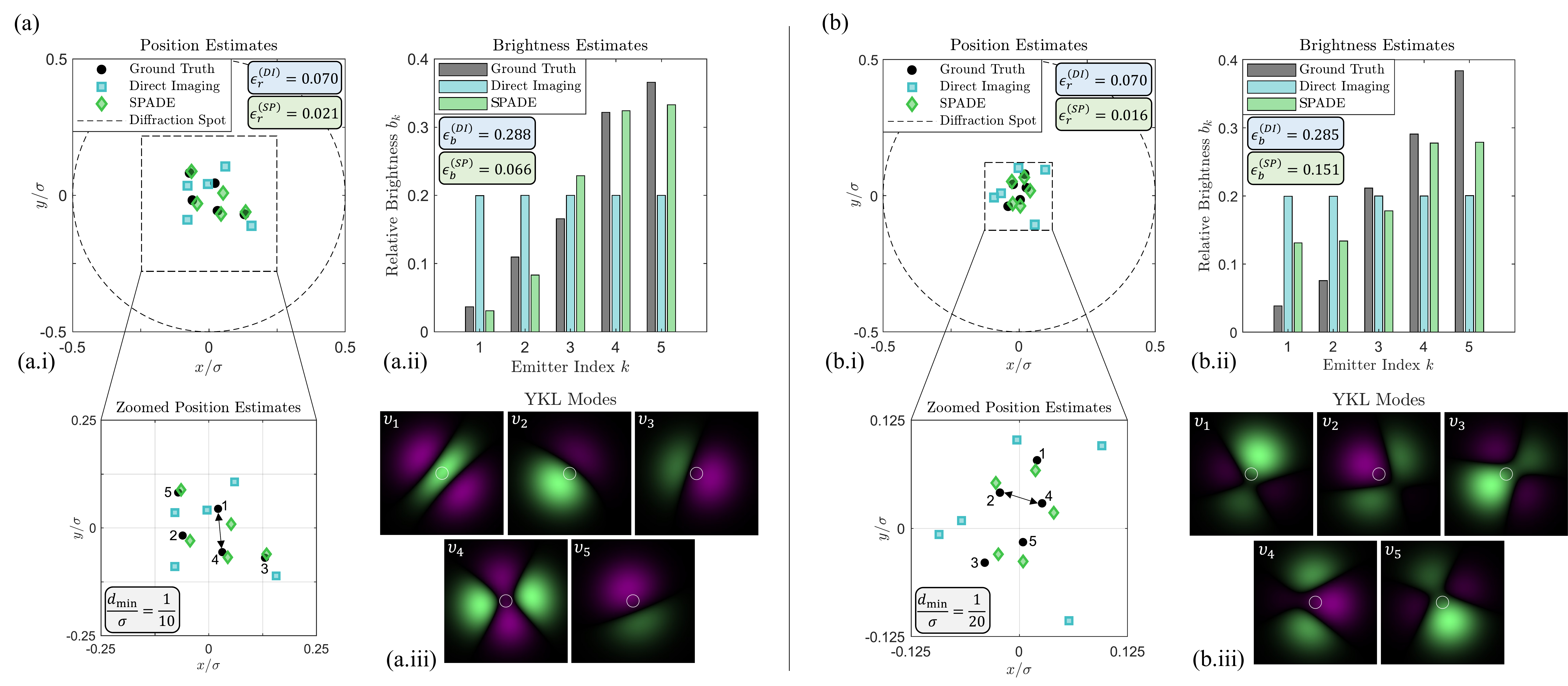}
\caption{Example estimates of sub-diffraction emitter positions and brightnesses made with our SPADE-enhanced protocol versus direct imaging. In either example, the emitter ensembles feature $K=5$ sources with a minimum pair-wise separation of \textbf{(a)} $d_{\text{min}} = \sigma/10$ and \textbf{(b)} $d_{\text{min}}=\sigma/20$. Sub-figures \textbf{(a-b.i)} show the emitter position estimates and localization error $\epsilon_{r}$ achieved during the calibration stage with our protocol (SPADE) versus direct imaging. Sub-figures \textbf{(a-b.ii)} show the emitter brightness estimates for SPADE and direct imaging. Note that direct imaging erroneously estimates uniform brightness from all emitters due to inaccurate localization and sub-optimal brightness sensitivity. In contrast, SPADE more accurately estimates the brightness of each emitter due to improved positional and brightness sensitivity afforded by the PAD measurement and the YKL measurement respectively. Sub-figures \textbf{(a-b.iii)} show the transverse optical field profile of the YKL modes at the image plane. White inset circles denote the sub-diffraction spot wherein the emitters reside. Note that the primary lobe for each YKL mode is approximately co-located with the emitter it seeks to discriminate. This maximizes the coupling between the $k^{\rm th}$ emitter and the $k^{\rm th}$ YKL mode under the constraint that the YKL modes remain mutually orthogonal. For these simulated examples, the number of photon resources used in the calibration and sensing stages was set to $M = N = 10^{7}$ with $M_1 = N_1 = 10^{6}$.} 
 \label{fig: Reconstructions}
\end{figure*}

\subsection*{Calibration Stage}

As demonstrated in \cite{Grace:2020}, the efficacy of spatial mode sorting for sub-diffraction localization is highly dependent on the alignment of the origin of the SPADE device with the geometric center of the emitter ensemble. To address this, we divide the calibration stage into two sequential measurements - direct imaging (DI) followed by spatial mode sorting in a PSF-adapted basis (PAD-SPADE). The photons captured with DI are used to estimate the geometric center of the scene. Subsequently, the PAD-SPADE instrument is aligned with this estimated center. This strategy aims to enhance the instrument's sensitivity to small off-axis emitter displacements by sieving out the dominant PSF mode component shared by all states $\ket{\psi(\vec{r}_k)}$. 

During this stage, the sample is optically pumped to generate steady-state fluorescence, ensuring that the sources emit with equal brightness. Therefore, throughout the calibration stage the parameter vector is given by,
\begin{equation}
    \bm{\theta}_0 \equiv [\mbf{r},\mbf{b}_0 ] \,,\qquad  \mbf{b}_0 \equiv \bigg[\frac{1}{K},\ldots,\frac{1}{K}\bigg]^{\intercal},
\end{equation}
where $\mbf{b}_0$ is taken to be known while $\mbf{r}$ remains unknown. Let $M$ be the total number of copies of state $\rho(\bm{\theta}_0)$ supplied during the calibration stage. We allocate $M_1$ copies to DI and the remaining $M_2 = M-M_1$ to PAD-SPADE.\footnote{The precise allocation of $M_1$ and $M_2$ is left unspecified. The optimal allocation in general depends on $\mbf{r}$ as shown in \cite{Grace:2020}, opening the possibility of potentially adaptive switching procedures between DI and PAD-SPADE.}

The direct imaging measurement is defined by the positive operator-valued measure (POVM) $\{\hat{\Pi}_{\vec{x}} = \dyad{\vec{x}} : \vec{x} \in \mathbb{R}^2\}$ comprised of the set of projectors on the position states. Let $\mbf{x} = [\vec{x}_1,\ldots,\vec{x}_{M_1}]$ be the collection of independent, identically distributed photon arrival positions observed across $M_1$ DI measurements. The joint probability density function for this measurement ensemble is,
\begin{equation}
    p(\mbf{x}|\mbf{r}) = \prod_{j=1}^{M_1} p(\vec{x}_j|\mbf{r}),
    \qquad  
    p(\vec{x}|\mbf{r}) = \Tr[\hat{\Pi}_{\vec{x}}\rho(\bm{\theta}_0)],
\end{equation}
where $p(\vec{x}|\mbf{r})$ is the probability of observing a single photon at the location $\vec{x}$ on the image plane given the emitter locations in their equal brightness arrangement. For sub-diffraction scenes, the maximum likelihood estimator for the geometric center is well approximated by the average arrival position,

\begin{equation}
\check{\vec{r}}_0 = \frac{1}{M_1}\sum_{j=1}^{M_1} \vec{x}_j.
\end{equation}
With this estimate in hand, we now consider the PAD-SPADE measurement which is defined by the POVM $\{\hat{\Psi}_{0}\} \cup \{\hat{\Psi}_{q} = \dyad{\psi_q} : q\in [1:Q]\}$
where $\psi_{1:Q}(\vec{x})$ are square-integrable orthonormal transverse modes whose origin is shifted to coincide with the estimated geometric midpoint. In particular, the first mode is defined to be the PSF mode shifted to the estimated geometric center $\psi_1(\vec{x}) \equiv \psi(\vec{x}-\check{\vec{r}}_0)$. The operator $\hat{\Psi}_0 = \hat{I}-\sum_{q=1}^{Q}\hat{\Psi}_{q}$ corresponds to a catch-all `bucket mode' that detects any photons lying in the orthogonal complement of $\text{span}\{\ket{\psi_{1:Q}}\}$. Let $\mbf{m} = [m_0,\ldots,m_{Q}]$ be the number of photons observed in each PAD mode across the SPADE measurement. Explicitly, $\mbf{m}$ is a multinomial random variable with distribution,
\begin{subequations}
\begin{align}
p(\mbf{m}|\mbf{r}) &= \text{Multinom}(\mbf{m}|\mbf{p}(\mbf{r}),M_2), \\
\mbf{p}(\mbf{r}) &= [p_0,\ldots,p_{Q}](\mbf{r}), \quad p_{q}(\mbf{r}) = \Tr[\hat{\Psi}_q \rho(\bm{\theta}_0)]. 
\end{align}
\end{subequations}
We subsequently calculate the maximum likelihood estimate for the source positions given by,
\begin{equation}
        \check{\mbf{r}} = \argmax_{\mbf{r}\in \mathbb{R}^{2K}} p(\mbf{m},\mbf{x}|\mbf{r}),
        \label{eqn: MLE Localization}
\end{equation}
where $p(\mbf{m},\mbf{x}|\mbf{r}) = p(\mbf{m}|\mbf{r})p(\mbf{x}|\mbf{r})$. Including the direct imaging measurements in the maximum likelihood estimate of $\check{\mbf{r}}$ serves to break degenerate solutions arising from symmetries in the PAD-SPADE modes. Indeed, the parametric family $p_q(\mbf{r})$ for SPADE is often not injective in $\mbf{r}$. That is, for any fixed set of emitter locations $\mbf{r}$, there exists a set of degenerate solutions $\{ \mbf{r}' \in \mathbb{R}^{K\times 2} : p_q(\mbf{r}') =p_q(\mbf{r})\,\, \forall\,\, q=1,\ldots,Q\}$ that give rise to identical SPADE statistics. Thus, introducing measurement diversity is critical for lifting these degeneracies.

\subsection*{Sensing Stage}
Having obtained an estimate of the source positions from Eqn. \ref{eqn: MLE Localization}, we transition to the sensing stage, which attempts to accurately estimate the brightness of the emitters at each $\gamma$. In particular, the parameter vector assumes the form,
\begin{equation}
\bm{\theta}_\gamma \equiv [\mbf{r},\mbf{b}_\gamma], \qquad \mbf{b}_\gamma \equiv \bigg[\frac{I_1^{(\gamma)}}{I_{0}^{(\gamma)}}, \ldots, \frac{I_{K}^{(\gamma)}}{I_0^{(\gamma)}}\bigg]^{\intercal}.
\end{equation}
Our goal is to determine the relative brightness parameters $\mbf{b}_{\gamma}$ given the estimate of the source positions. To do so, the sensing stage administers DI measurements followed by spatial mode sorting in the Yuen-Kennedy-Lax basis (YKL-SPADE). Let $N^{(\gamma)}$ be the number of copies of state $\rho(\bm{\theta}_{\gamma})$ available. As in the calibration stage, we allocate $N^{(\gamma)}_{1}$ to DI measurements and $N^{(\gamma)}_{2} = N^{(\gamma)}-N^{(\gamma)}_{1}$ to YKL-SPADE measurements.\footnote{Again, the division of photon resources $N_1^{(\gamma)}$ and $N_2^{(\gamma)}$ within the sensing stage is left unspecified. However, inspired by the asymptotic optimality proof of \cite{Hayashi:2005, GongBash:2024} we suggest the allocation $N_1 = f(N)$ where $f(\cdot)$ satisfies $\lim_{N\rightarrow \infty} f(N) \rightarrow\infty$ and $\lim_{N\rightarrow\infty} \frac{f(N)}{N}=0$. A common choice is $f(N)=\lceil \sqrt{N}\,\rceil$ such that $N_1 = \lceil \sqrt{N} \rceil$.}
Let the collection of outcomes of the DI measurements at a particular $\gamma$ be $\mbf{x}_{\gamma} = [\vec{x}^{(\gamma)}_{1},\ldots,\vec{x}^{(\gamma)}_{N_1^{(\gamma)}}]$. These samples are used to formulate a brightness pre-estimate $\check{\mbf{a}}_{\gamma}$ given by,
\begin{subequations}
\begin{align}
    \check{\mbf{a}}_{\gamma} &= \argmax_{\mbf{b}\in \mathcal{S}_K} p(\mbf{x}_{\gamma}|\check{\mbf{r}},\mbf{b}),\, {\text{with}}\\
    p(\mbf{x}_{\gamma}|\check{\mbf{r}},\mbf{b}) &= \prod_{j=1}^{N_1^{(\gamma)}}p(\vec{x}_j^{(\gamma)}|\check{\mbf{r}},\mbf{b}), \, \, p(\vec{x}|\check{\mbf{r}},\mbf{b}) = \Tr[\hat{\Pi}_{\vec{x}}\rho(\check{\mbf{r}},\mbf{b})]. 
\end{align}     
\end{subequations}

\begin{figure*}
 \centering
 \includegraphics[width=\linewidth]{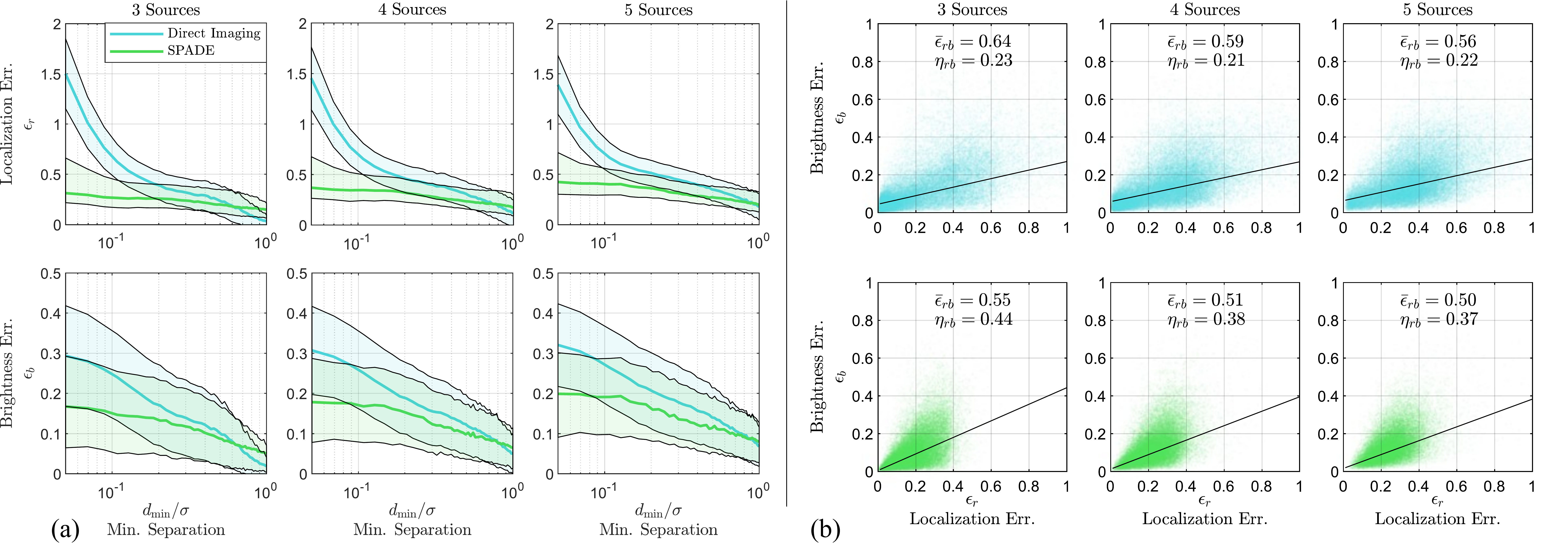}
 \caption{Monte-Carlo statistics demonstrating the relative localization error and brightness estimation error improvement achieved with our two-stage protocol compared to direct imaging. \textbf{(a)} The localization error $\epsilon_r$ (top) and brightness estimation error $\epsilon_b$ (bottom) as a function of the minimum pair-wise separation $d_{\text{min}}$ of emitters in the scene. The shaded regions delineate the standard deviation of the mean error computed from the Monte-Carlo simulations. Our protocol exhibits a pronounced advantage over direct imaging in the deep sub-diffraction regime where $d_{\text{min}}<\sigma/8$. In this regime our protocol improves the average localization accuracy by $\sim 6\times$ and the average brightness estimation accuracy by $\sim 2\times$. \textbf{(b)} Statistical correlation between localization error and brightness estimation error for Direct Imaging (top) and our protocol (bottom) quantified by the Pearson correlation coefficient $\bar{\epsilon}_{rb}$. The slopes $\eta_{rb}$ of the linear data fits quantify how sensitive the brightness error is to the localization error. All simulations were run using $M = N = 10^{5}$ photons with $M_1 = N_1 = 10^{4}$. The PAD-SPADE receiver was configured to sort the first $66$ Hermite-Gauss modes for which the mode indices $n+m\leq 10$.  }
 \label{fig: K-Source Monte-Carlos}
\end{figure*}

Subsequently, we use the brightness pre-estimate to define the YKL-SPADE measurement. This measurement consists of a set of orthogonal projectors that minimize the probability of error for discriminating the states $\ket{\psi(\check{\vec{r}}_k)}$ with priors $\check{a}_k^{(\gamma)}$ \cite{YKL:1975},
\begin{equation}
P_{\text{e,min}}(\check{\mbf{r}},\check{\mbf{a}}_{\gamma}) = \min_{\{\ket{\upsilon_k}\}_{k=1}^{K}} 1- \sum_{k=1}^{K} \check{a}_k^{(\gamma)} |\braket{\upsilon_k}{\psi(\check{\vec{r}}_k)}|^2.
\label{eqn: Min Prob Error}
\end{equation}
The complete POVM associated with the YKL-SPADE measurement is defined as $\{ \hat{\Upsilon}^{(\gamma)}_{0} \} \cup \{\hat{\Upsilon}^{(\gamma)}_{k} = |\upsilon_k^{(\gamma)}\rangle \langle \upsilon_k^{(\gamma)}| : k\in[1:K]\}$ where $|\upsilon_k^{(\gamma)}\rangle$ are orthonormal states minimizing the error in Eqn. \ref{eqn: Min Prob Error} and $\hat{\Upsilon}^{(\gamma)}_{0} = \hat{I}-\sum_{k=1}^{K}\hat{\Upsilon}^{(\gamma)}_k$ is yet another `bucket mode' that captures the orthogonal complement of $\text{span}\{\ket{\psi(\check{\vec{r}}_k)}\}_{k=1}^{K}$. Importantly, we note that the span of the estimated states $\ket{\psi(\check{\vec{r}}_k)}$ may not necessarily be equal to the span of the true states $\ket{\psi(\vec{r}_k)}$ due to residual errors in the emitter localization performed during the calibration stage. Therefore, the YKL-SPADE measurement may project onto a subspace that does not contain all of the true states, which requires introducing the bucket operator $\hat{\Upsilon}_0$. Additionally, if the brightness pre-estimates are poor, the efficacy of the YKL measurement deteriorates as the priors for the photon emission probability of each emitter are mismatched to the true underlying regularity with which they produce a photon. While the parameter-dependence of the YKL-SPADE measurement stands to provide a highly sensitive brightness measurement optimized to the exact emitter configuration, it pays a penalty if the pre-estimates of the parameters are inaccurate.

Let $\mbf{n}_{\gamma}=[n_0^{(\gamma)},\ldots,n_{K}^{(\gamma)}]$ be the number of photons observed in each YKL mode such that $\mbf{n}$ is a multinomial random variable with distribution:
\begin{subequations}
\begin{align}
    p(\mbf{n}_{\gamma}|\bm{\theta}) &= \text{Multinom}(\mbf{n}_{\gamma}|\mbf{q}_{\gamma}(\bm{\theta}),N_2^{(\gamma)}),\\
    \mbf{q}_{\gamma}(\bm{\theta}) &= [q_0^{(\gamma)},\ldots,q_{K}^{(\gamma)}](\bm{\theta}), \quad q_{k}^{(\gamma)}(\bm{\theta}) = \Tr[\hat{\Upsilon}^{(\gamma)}_{k}\rho(\bm{\theta})].  
\end{align}
\end{subequations}
For each $\gamma$ we estimate the brightness via,
\begin{equation}
\check{\mbf{b}}_{\gamma} = \argmax_{\mbf{b}\in \mathcal{S}_{K}} p(\mbf{n}_{\gamma},\mbf{x}_{\gamma}|\check{\mbf{r}},\mbf{b}),
\end{equation}
where $p(\mbf{n}_{\gamma},\mbf{x}_{\gamma}|\check{\mbf{r}},\mbf{b}) = p(\mbf{n}_{\gamma}|\check{\mbf{r}},\mbf{b})p(\mbf{x}_{\gamma}|\check{\mbf{r}},\mbf{b})$. Combining the estimates from the calibration and sensing stages, our protocol returns an estimate of all parameters $\check{\bm{\theta}}_{\gamma} = [\check{\mbf{r}},\check{\mbf{b}}_{\gamma}]$.

\section*{Results}
\label{sec: Results}
We evaluate the statistical performance improvement that our SPADE-enhanced protocol offers over direct imaging for sub-diffraction ensembles consisting of $K=3,4,5$ emitters. Specifically, the benchmark direct imaging protocol is equivalent to our SPADE-enhanced protocol, but with $M_1 = M$ and $N_1^{(\gamma)}= N^{(\gamma)}$ so that no photons are allocated to the SPADE measurements (i.e. $M_2 = 0$ and $N_2^{(\gamma)} = 0$). In this way, both the direct imaging protocol and our SPADE-enhanced proposal receive the same set of quantum states - any divergence in their performance then depends solely on how these states are measured. 

Throughout our numerical analysis, we assume a linear shift-invariant imaging system with a Gaussian PSF $\psi(\vec{x}) = \exp(-(x^2 +y^2)/4\sigma^2)/\sqrt{2\pi \sigma^2}$ that reasonably approximates the Bessel PSF corresponding to canonical imaging systems with hard circular apertures. Under this choice of PSF, the Hermite-Gauss modes constitute a natural choice for the PAD-SPADE measurement.\footnote{The Hermite-Gauss modes are defined as,
\begin{equation}
    \psi_{nm}(\vec{x}) = \frac{H_n\left(\frac{x}{\sigma\sqrt{2}}\right)}{\sqrt{2^{n}n!}}\frac{H_m\left(\frac{y}{\sigma \sqrt{2}}\right)}{\sqrt{2^{m}m!}}\psi(\vec{x}),
\end{equation}
where $H_n(x)$ are the physicist's Hermite polynomials and $n,m \in [0:\infty]$.} Figures \ref{fig: Reconstructions}(a.i) and \ref{fig: Reconstructions}(b.i) exemplify the localization accuracy achieved using our protocol compared to direct imaging for collections of five emitters whose minimum pairwise separation is $\sigma/10$ and $\sigma/20$, respectively. For state-of-the-art confocal microscopes, such distances correspond to $10$'s of nanometers. Figures \ref{fig: Reconstructions}(a.iii) and \ref{fig: Reconstructions}(b.iii) highlight the improved brightness estimation accuracy achieved with our protocol. This improvement arises from the accurate localization estimates afforded by the PAD-SPADE measurements combined with the information efficiency of the YKL-SPADE measurement. Figures \ref{fig: Reconstructions}(a.iv) and \ref{fig: Reconstructions}(b.iv) show the YKL modes associated with estimated positions $\check{\mbf{r}}$ and the preliminary brightness estimate $\check{\mbf{a}}$ obtained from the previous measurements of the protocol. Intuitively, the primary lobes of the $k^{\rm th}$ YKL mode appear to spatially coincide with the orientation of the $k^{\rm th}$ emitter relative to the rest of the ensemble. These modes strike a balance between coupling as much light as possible from their target emitter while remaining orthogonal.

To statistically quantify the performance advantage of our protocol, we ran Monte Carlo simulations on randomly generated scenes with a minimum pairwise separation $d_{\text{min}}(\mbf{r}) =\min_{j\neq k} |\vec{r}_j - \vec{r}_k|$; the length scale corresponding to the `hardest-to-resolve' source pair.  For each Monte-Carlo scene sample, we evaluated the localization error $\epsilon_{r}$ and the brightness error $\epsilon_{b}$ given by,
\begin{subequations}
\begin{align}
    \epsilon_{r}(\mbf{r},\check{\mbf{r}}) &= \frac{1}{K}\sum_{k=1}^{K} \frac{|\vec{r}_k - \check{\vec{r}}_k|}{d_{\text{min}}(\mbf{r})},\\
    \epsilon_{b}(\mbf{b},\check{\mbf{b}}) &= \frac{1}{2}\sum_{k=1}^{K} |b_k - \check{b}_k|.
\end{align}
\label{eqn: Error Metrics}
\end{subequations}
The localization error metric $\epsilon_r$ is the average Euclidean distance between an estimated emitter location and its ground-truth position normalized by the minimum pair-wise separation in the scene. For sub-diffraction constellations, this dimensionless error metric remains agnostic to both the scale of the scene and the number of emitters. Meanwhile, the brightness error metric $\epsilon_{b} \in [0,1]$ is the so-called `total-variation' commonly used to quantify the statistical distance between two discrete probability distributions, in this case $\mbf{b}$ and $\check{\mbf{b}}$. This metric is also intrinsically agnostic to the number of emitters.\footnote{Note that the ordering of the emitter indices matters here. If we were to permute the indices of the estimates (which is equivalent to reassigning each emitter to a different ground truth) then the error metrics would change. Thus in the definition of Equations \ref{eqn: Error Metrics} we implicitly assume an ordering of the parameters that minimizes the localization error:
\begin{equation}
    \check{\bm{\theta}} \leftarrow R \check{\bm{\theta}} \quad : \quad R = \argmin_{R\in \mathcal{P}_{K}} \epsilon_r(\mbf{r},R\check{\mbf{r}})
\end{equation}
where $\mathcal{P}_{K}$ is the permutation group of $K\times K$ matrices.} 

\begin{figure*}
\includegraphics[width=.9\linewidth]{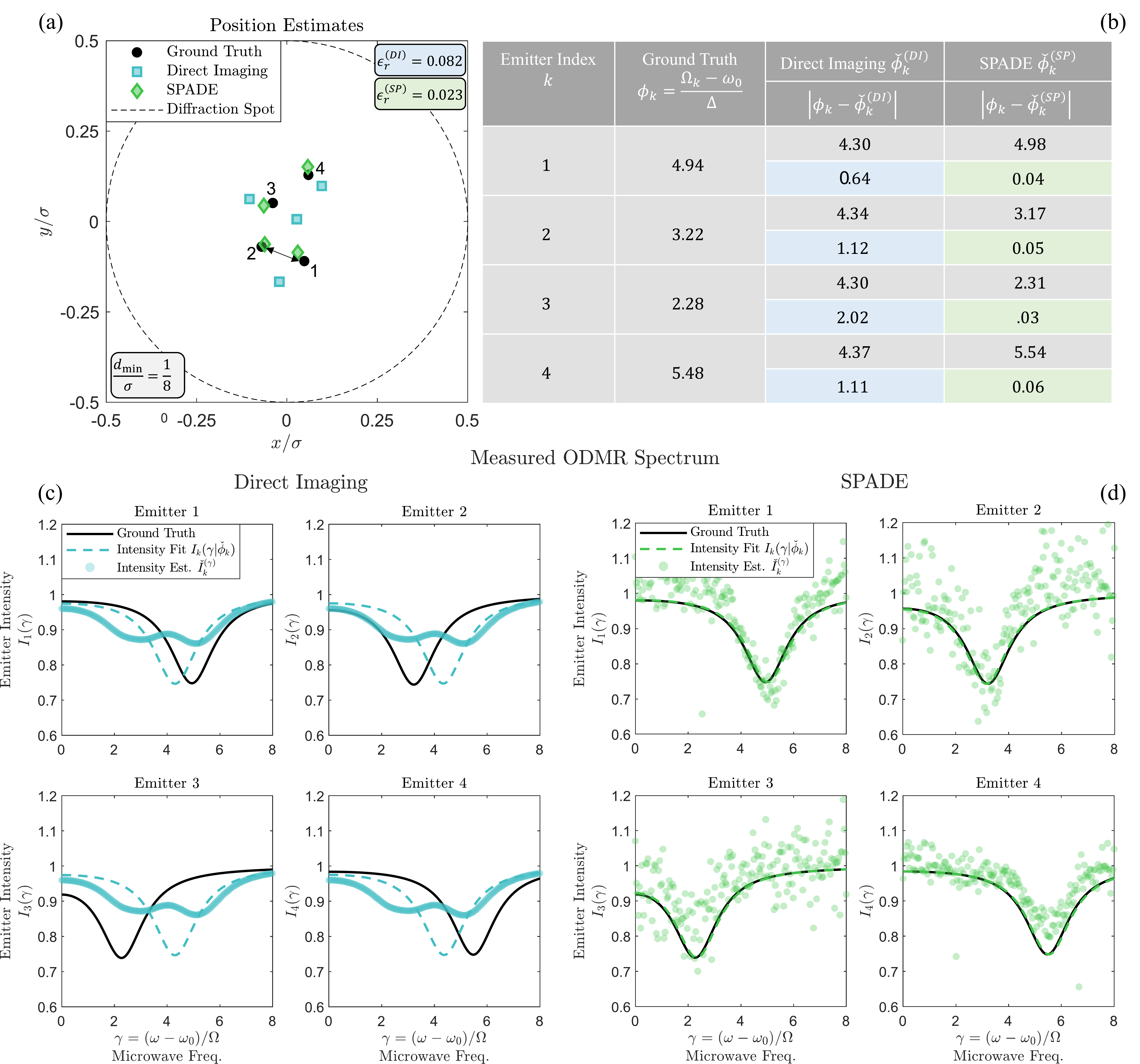}
 \caption{Simulated CW ODMR experiment involving four sub-diffraction NV centers with minimum pairwise separation $d_{\text{min}} = \sigma/8$ and non-radiative on-resonance transition probability $\chi=0.5$. Here, the experimental modulation variable is the dimensionless centered microwave frequency $\gamma = (\omega -\omega_0)/\Delta$. \textbf{(a)} Shows the estimated emitter positions made with our protocol as compared to direct imaging ($\sim 4\times$ improvement in localization accuracy). Table \textbf{(b)} shows the value of the target field (in this case the dimensionless Zeeman frequency $\phi_k = \frac{\Omega_k - \omega_0}{\Delta} \propto B_{k}$ encoding local magnetic field strength) recovered from a least-squares fit of the parametric photoluminescence model $I(\gamma|\phi)$ in Eqn. \ref{eqn: CW ODMR Model}
to the estimated emitter intensities $\check{I}_{k}^{(\gamma)}= N_2^{(\gamma)}\check{b}_{k}^{(\gamma)}$. The highlighted table entries compare the error in the estimated field value $|\check{\phi}_{k}-\phi_k|$ for direct imaging (blue) and our protocol (green). We observe an average improvement of $28.5 = [\sum_{k=1}^{4}(\phi_{k}-\check{\phi}_{k}^{({\rm DI})})^2/\sum_{k=1}^{4}(\phi_{k}-\check{\phi}_{k}^{({\rm SP})})^2]^{1/2}$ in root-mean-squared-error of the Zeeman frequency estimates achieved with our protocol over direct imaging. Sub-figures \textbf{(c)} and \textbf{(d)} show the series of intensity estimates $\check{I}_{k}^{(\gamma)}$ made with our protocol versus direct imaging for each emitter over the course of the ODMR microwave frequency sweep. While our protocol exhibits greater variance over the estimated emitter intensities, direct imaging suffers from severe estimator bias resulting in lower fidelity estimates of the target field parameter $\phi$. For these simulations we used photon allocations $M=N^{(\gamma)} = 10^7$ and $M_1=N_1^{(\gamma)} = 10^6$ in the calibration and sensing stages.}
 \label{fig: CW ODMR}    
\end{figure*}

Figure \ref{fig: K-Source Monte-Carlos}(a) depicts the statistical mean of the localization error $\epsilon_r$ and brightness error $\epsilon_b$ as a function of the minimum pair-wise emitter spacing $d_{\text{min}}$. In the deep sub-diffraction regime where $d_{\text{min}} \ll \sigma$ and all sources are emitters are tightly concentrated, our protocol outperforms direct imaging in localization accuracy by $\sim 6\times$. Furthermore, while the localization error for direct imaging appears to blow up as $d_{\text{min}} \rightarrow 0$, the localization error for our protocol remains stable in the same limit.  

Figure \ref{fig: K-Source Monte-Carlos}(b) illustrates the statistical correlation between $\epsilon_{r}$ and $\epsilon_{b}$ quantified via the Pearson correlation coefficient $\bar{\epsilon}_{rb}$.\footnote{The Pearson correlation coefficient is given by,
\begin{equation}
\bar{\epsilon}_{rb} \equiv \frac{\sum_{i=1}^{N_s} ({\epsilon}_r^{(i)} - \bar{{\epsilon}}_r)(\epsilon_b^{(i)}-\bar{\epsilon}_b)}{\sqrt{\sum_{i=1}^{N_s} ({\epsilon}_r^{(i)} - \bar{{\epsilon}}_r)^2}\sqrt{\sum_{j=1}^{N_s} (\epsilon_b^{(j)}-\bar{\epsilon}_b)^2}} 
\label{eqn: Pearson Correlation Coeff}
\end{equation}
where $N_s$ is the total number of monte-carlo trials compiled while $\bar{{\epsilon}}_r = \frac{1}{N_s}\sum_{i=1}^{N_s} {\epsilon}_{r}^{(i)}$ and $\bar{\epsilon}_b = \frac{1}{N_s}\sum_{i=1}^{N_s} \epsilon_{b}^{(i)}$ denote the empirical mean of the relative localization error and brightness error respectively.} This quantity is $\pm 1$ if the localization and brightness error samples lie along a line (perfectly correlated/anti-correlated) and $0$ if the are isometrically distributed (no correlation). Both our protocol and direct imaging exhibit positive correlations in the range $0.5 \leq \bar{\epsilon}_{rb} \leq 0.65$, indicating that improved localization accuracy should improve brightness estimation accuracy. This is to be expected since the brightness estimates are conditioned on the emitter location estimates. Thus any error in the emitter localization translates to a bias in the brightness estimates. While the correlation coefficient quantifies the statistical dependence between the brightness error and the localization error, it does not capture how quickly $\epsilon_b$ is expected to grow with respect to $\epsilon_r$. Therefore, Fig. \ref{fig: K-Source Monte-Carlos}(b) also contains linear fits to the Monte-Carlo errors. The slope of this fit $\eta_{rb} \approx \frac{d\epsilon_b}{d\epsilon_r}$ approximately measures how sensitive the brightness error is to the localization error. We observe that the errors $\epsilon_r$ and $\epsilon_b$ are less correlated in our protocol compared to direct imaging. However, the brightness error is substantially more sensitive to localization error in our protocol compared to direct imaging presumably because the design of the YKL measurement depends on the emitter location estimates made in the calibration stage. If the location estimates are erroneous, then we expect the measurement to be very insensitive to brightness because a larger fraction of photons will appear in the bucket mode $\hat{\Upsilon}_0$, which contains limited information.

\begin{figure*}
    \includegraphics[width=.9\linewidth]{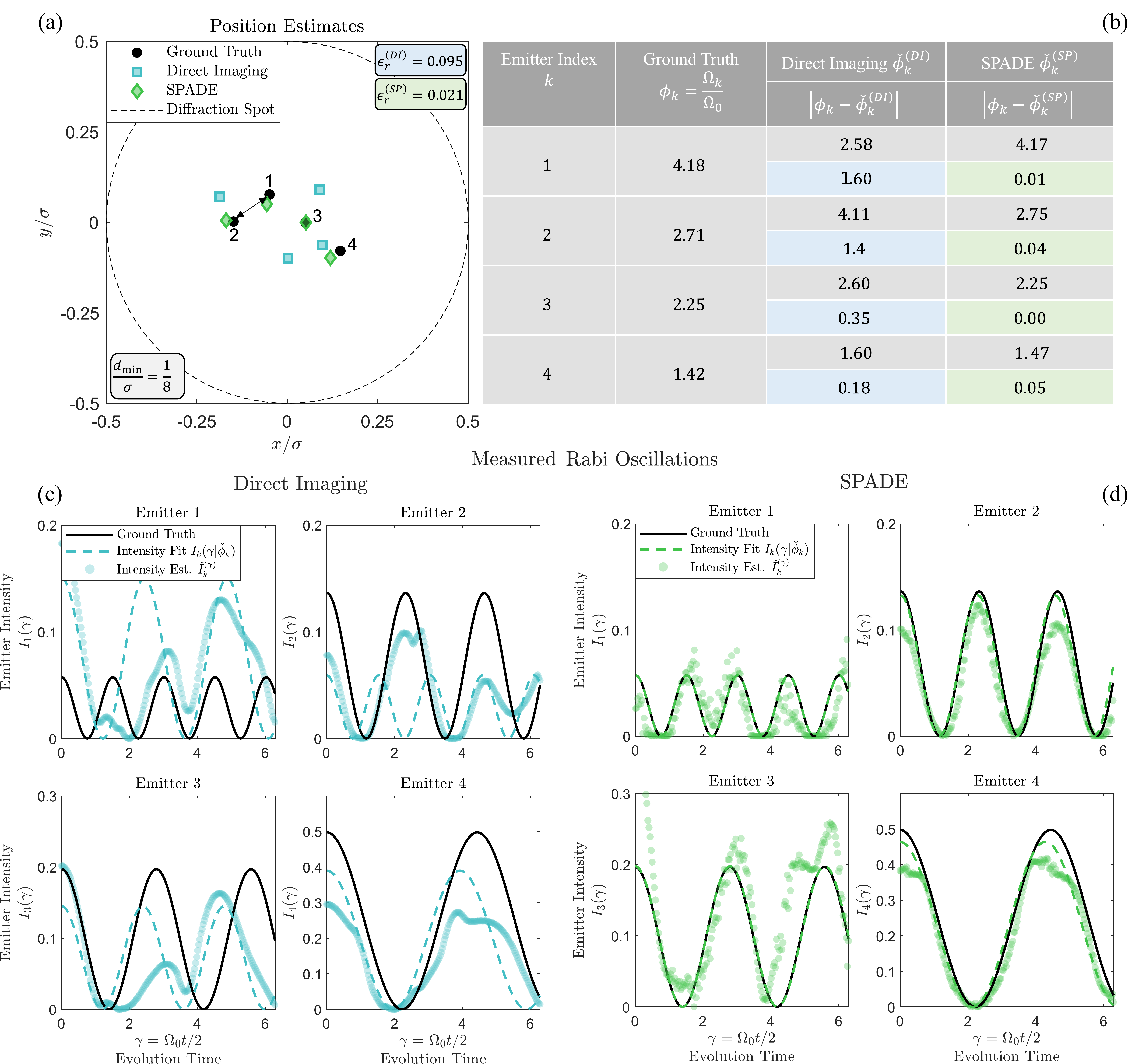}
 \caption{Simulated Rabi experiment involving four sub-diffraction NV centers with minimum pairwise separation $d_{\text{min}}=\sigma/8$. Here, the experimental modulation variable is the dimensionless evolution time $\gamma = \Omega_0 t/2$ of the emitters where $\Omega_0$ is the resonant Rabi frequency. \textbf{(a)} Shows the estimated emitter positions achieved with our protocol as compared to direct imaging ($\sim 5\times$ improvement in localization accuracy). Table \textbf{(b)} shows the value of the target field (the Rabi frequency of each emitter $\phi_k = \frac{\Omega_k}{\Omega_0}$) recovered from a least-squares fit of the parametric photoluminescence model $I(\gamma|\phi)$ of Eqn. \ref{eqn: Rabi Model} to the collection of estimated emitter intensities $\check{I}_{k}^{(\gamma)} = N_{2}^{(\gamma)}\check{b}_{k}^{(\gamma)}$. The highlighted table entries compare the error in the estimated field value $|\check{\phi}_{k}-\phi_k|$ for direct imaging (blue) and our protocol (green). We observe an average improvement of $33.4 = [\sum_{k=1}^{4}(\phi_{k}-\check{\phi}_{k}^{({\rm DI})})^2/\sum_{k=1}^{4}(\phi_{k}-\check{\phi}_{k}^{({\rm SP})})^2]^{1/2}$ in root-mean-squared-error of the Rabi frequency estimates achieved with our protocol over direct imaging.  Sub-figures \textbf{(c)} and \textbf{(d)} show the series of intensity estimates $\check{I}_{k}^{(\gamma)}$ made with our protocol versus direct imaging for each emitter over the course of the state evolution time. For these simulations we used photon allocations $M=N^{(\gamma)} = 10^7$ and $M_1=N_1^{(\gamma)} = 10^6$ in the calibration and sensing stages. }
 \label{fig: Rabi Oscillation}
\end{figure*}

\subsection*{Continuous Wave ODMR}
\label{subsec: CW-ODMR}

 Continuous Wave ODMR aims to measure $\phi(\vec{x}) = \vec{B}(\vec{x}) \cdot \vec{n}$, or the component of a stationary magnetic field along an arbitrary crystal axis $\vec{n}$ of the NV centers. As an example of our protocol in action, we consider a simple ODMR experiment consisting of $K=4$ sub-diffraction NV centers with minimum pair-wise separation $d_{\text{min}}=\sigma/8$ shown in Fig. \ref{fig: CW ODMR}. For simplicity, we assume all NV axes in the ensemble are oriented along the same direction. The local magnetic field component at each NV is $B_k = \vec{B}(\vec{r}_k)\cdot \vec{n}$. The intensity model for each emitter is given by,
\begin{subequations}
    \begin{align}
        I_k(\omega|\Omega_k) &\propto 1 - \frac{\chi}{2}\big( L(\omega,-\Omega_k) + L(\omega,+\Omega_k)\big),\\
        L(\omega,\Omega) &= \frac{1}{1+(\omega-\Omega_0-\Omega)^2/\Delta^2}.
    \end{align}
    \label{eqn: CW ODMR Model}
\end{subequations}
where the experimentally tunable independent variable is the frequency of an external microwave drive field $\omega$. Additionally, $\Omega_k$ is the detuning frequency of the Zeeman splitting energy induced by the local magnetic field strength at each emitter, while $\chi\in(0,1)$ is the transition probability to the vibrational non-radiative triplet state of the NV, and $L(\cdot)$ is a dimensionless Lorentzian response function. The Lorentzian involves the zero-field frequency $\Omega_0$ and the response linewidth $\Delta$. The detuning frequency $\Omega_k$ encodes the local magnetic field magnitude $B_k$ along the NV axis via, $\Omega_k = \sqrt{\big(g \mu_B  B_k/\hbar \big)^2 + E^2}$ where $g\approx 2.0$ is the g-factor, $\mu_B$ is the Bohr magneton, $\hbar$ is the reduced Planck constant, and $E$ is the off-axis zero-field splitting factor induced by strain in the diamond lattice. 

In Figure \ref{fig: CW ODMR}(a) we compare the estimated locations of the ensemble acquired during the calibration stage for our proposed protocol and direct imaging. Figures \ref{fig: CW ODMR}(c-d) showcase the brightness estimates made over the course of a simulated microwave frequency sweep which we then use to estimate the Zeeman frequency $\Omega_k$ via a least-squares best-fit. Note that the brightness estimates made with our protocol exhibit higher variance compared to the brightness estimates of direct imaging, however, the bias of the brightness estimates made with our protocol nearly zero. As a consequence, Fig. \ref{fig: CW ODMR}(b) shows that our protocol recovers estimates of the Zeeman frequency at each emitter which are consistently more precise than those attained with direct imaging alone.

\subsection*{Rabi Oscillations}
Our second example considers monitoring the Rabi oscillations of $K=4$ sub-diffraction emitters with $d_{\text{min}}=\sigma/8$ shown in Fig. \ref{fig: Rabi Oscillation}. While not an AC sensing protocol on its own, a Rabi measurement usually serves as an initial calibration step for an AC NV experiment. Thus, the ability to resolve individual Rabi signals is indicative of expected performance during a more complex AC measurement protocol. In this example, we model each emitter as a two-level atomic system in the presence of a periodic drive that couples the excited state and the ground state. Each emitter is assumed to have a Rabi frequency $\Omega_k = \sqrt{\Omega_0^2 + \Delta_k^2}$ that depends on the local detuning $\Delta_k$ at the emitter position. The resonant Rabi frequency $\Omega_0$ is assumed to be the same for all emitters. We further assume that each emitter is prepared in the excited state at time $t=0$. Each emitter will radiate over time in accordance with the Rabi model,
\begin{equation}
I_k(t|\Omega_k)\propto
\bigg(\frac{\Omega_0}{\Omega_k}\bigg)^2 \cos^2\bigg(\frac{\Omega_k t}{2} \bigg).
\label{eqn: Rabi Model}
\end{equation}
We wish to ascertain the individual frequencies $\Omega_k$ by curve-fitting the measured brightness as a function of time to the Rabi model. In this case, the independent variable $\gamma$ in the experiment is the evolution time $t$ of the two-level system. Fig. \ref{fig: Rabi Oscillation}(a) shows the emitter localizations achieved with our protocol and direct imaging. Figures \ref{fig: Rabi Oscillation}(c-d) show the brightness values estimated for each value of $\gamma$ along with the least-squares best-fit curve under the Rabi model. The quality of the Rabi frequency estimates is summarized in the table of Fig. \ref{fig: Rabi Oscillation}(b).

\section*{Discussion}
Here we offer an information-theoretic analysis of the YKL-SPADE measurement in order to elucidate its utility for brightness estimation. While sub-diffraction emitter localization (and the advantages of the PAD-SPADE basis) has been studied extensively over the past decade, the task of brightness estimation has received considerably less attention. For this reason, we primarily focus on the theoretical performance of the YKL-SPADE measurement, for which a more comprehensive discussion can be found in Sections I and III of the Supplementary Information. The curious reader is referred to \cite{Gessner:2023,Boeschoten:2026} for a general quantum treatment of mode-encoded parameter estimation which encapsulates emitter localization.

\begin{figure*}
 \centering
\includegraphics[width=\linewidth]{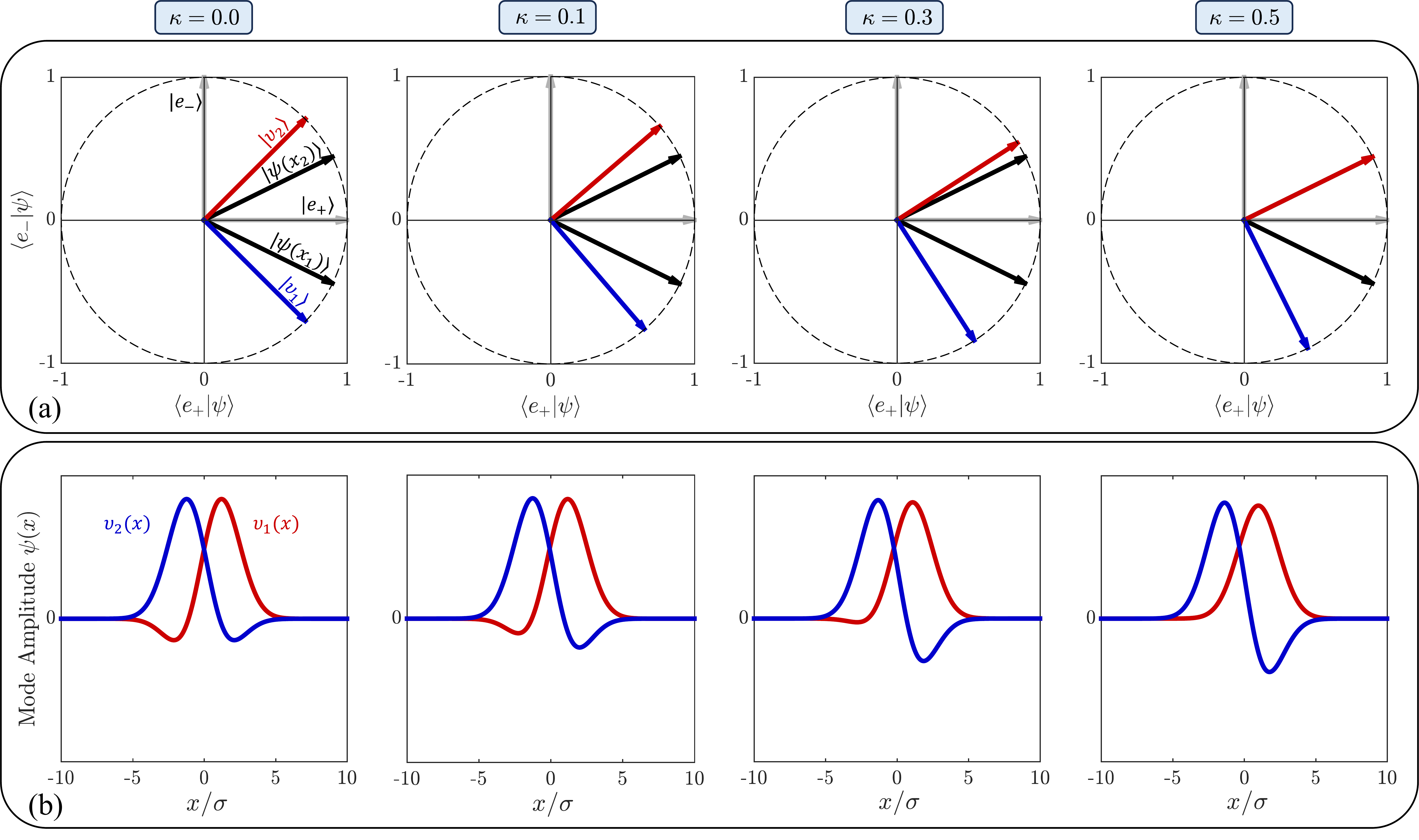}
 \caption{Geometry of the SLD/YKL eigenstates $\ket{\upsilon_1}$ and $\ket{\upsilon_2}$ for optimal brightness estimation between two point sources. The underlying source separation is fixed at $s=\sigma$ while each sub-panel depicts a different value of the brightness bias $\kappa$ transitioning from the balanced brightness case to the unbalanced case. \textbf{(a)} Depicts the projector directions in the $2D$ subspace formed by the span of $\ket{\psi(x_1)}$ and $\ket{\psi(x_2)}$. As brightness bias increases, the projectors rotate to increase classification accuracy of the brighter source at the expense of increasing classification accuracy of the dimmer source. \textbf{(b)} Depicts the spatial modes at the image plane associated with the optimal projective measurements $\upsilon_1(x)$ and $\upsilon_2(x)$ assuming a gaussian PSF.}
 \label{fig: Two-Source Brightness POVM}
\end{figure*}

\subsection*{Analysis of $K=2$ Emitters: \\The SLD-YKL Correspondence}
For a pedagogical introduction, we first restrict our attention to a system of two incoherent emitters separated along one dimension - later we generalize to multiple emitters. The positions of the two emitters are given by $\mbf{r}=[x_1,x_2] \in \mathbb{R}^{2}$ with relative brightness parameters $\mbf{b}=[b_1, b_2] \in S_2$ such that  $b_1 + b_2 = 1$. The single-photon state of the field is then expressed as,
\begin{equation}
\rho(\bm{\theta}) =b_{1} \dyad{\psi(x_1)} + b_{2}\dyad{\psi(x_2)}
\label{eqn: 2-Source Density}
\end{equation}
The operational meaning of the mixed state in Eq. \ref{eqn: 2-Source Density} can be described in the following way. Each photon arriving at the image plane is definitively in a pure state, either $\ket{\psi(x_1)}$ or $\ket{\psi(x_2)}$. The supplied state is generated randomly by a classical process with probability $b_1$ and $b_2=1-b_1$ (e.g. a biased coin flip). The mixed state $\rho(\bm{\theta})$ simply describes our classical ignorance about which state was generated. We may therefore imagine a (random) binary sequence of $N$ states which are the true states we receive. For example, one such sequence may look like,
$$
\underbrace{
\ket{\psi(x_1)},\ket{\psi(x_2)},\ket{\psi(x_2)},\ket{\psi(x_1)},\ldots,\ket{\psi(x_1)}}_{N \text{ state sequence}}.
$$
In the limit $N\rightarrow\infty$, the relative frequency with which $\ket{\psi(x_1)}$ ($\ket{\psi(x_2)}$) appears approaches $b_1$ ($b_2$). Therefore, brightness estimation appears tantamount to identifying the relative frequency with which each state appeared in the sequence. From this perspective, we may intuit that the optimal measurement for determining the brightness should relate to the optimal hypothesis testing measurement for discriminating $\ket{\psi(x_1)}$ and $\ket{\psi(x_2)}$. 

Since the brightness parameters are subject to a constraint equation, let $\kappa=(b_2-b_1)/2 \in (-\frac{1}{2},\frac{1}{2})$ be a single free parameter called the {\em brightness bias} of the sources. Indeed we will prove that the optimal measurement for estimating the brightness bias $\kappa$, given by projectors onto the eigenstates of the SLD $\hat{L}_{\kappa}$, is equivalent to the YKL measurement for binary state discrimination. This measurement is also known as the Helstrom measurement in the special case of binary hypothesis testing \cite{Helstrom:1976}.

We begin by calculating the SLD of the brightness parameter $\kappa$ which satisfies the Lyapunov equation, 
\begin{equation}
    \partial_{\kappa} \rho(\bm{\theta})= \hat{L}_{\kappa}\circ \rho(\bm{\theta}).
\end{equation}
Then define the operators $\hat{\Delta}_{\pm} \equiv \dyad{\psi(x_2)} \pm \dyad{\psi(x_1)}$ such that $\rho(\bm{\theta}) = \frac{1}{2}\hat{\Delta}_{+} + \kappa \hat{\Delta}_{-}$ and the Lyapunov equation reduces to,
\begin{equation}
\hat{\Delta}_{-}=\hat{L}_{\kappa}\circ \rho. 
\label{eqn: Lyap Brightness Param}
\end{equation}
Defining the overlap $\varphi = \braket{\psi(x_1)}{\psi(x_2)}\in \mathbb{R}$ (assuming a real-valued PSF) and the ordered orthonormal basis $\{\ket{e_{+}},\ket{e_{-}}\}$,
$$
\ket{e_{\pm}} = \frac{1}{\sqrt{2(1\pm \varphi)}}(\ket{\psi(x_2)}\pm \ket{\psi(x_1)}),
$$
the operators $\hat{\Delta}_{\pm}$ can be written as matrices,
\begin{equation}
\Delta_{+} = \begin{bmatrix}
    (1+\varphi)  & 0 \\
    0 & (1-\varphi)
\end{bmatrix},\,\, \Delta_{-} = \begin{bmatrix}
    0 & \sqrt{1-\varphi^2}\\
    \sqrt{1-\varphi^2} &0
\end{bmatrix}.
\end{equation}
One can verify that the solution to the Lyapunov equation of \ref{eqn: Lyap Brightness Param} is given by the SLD matrix,
\begin{equation}
L_{\kappa} \equiv \frac{2}{\tau}
\begin{bmatrix}
 -2\kappa(1-\varphi) & \sqrt{1-\varphi^2}\\
 \sqrt{1-\varphi^2} & -2\kappa(1+\varphi)
\end{bmatrix}, 
\end{equation}
where $\tau \equiv (1-(2\kappa)^2)$. With the SLD in hand, we now turn to the YKL measurement which admits a closed-form expression in the case of binary hypothesis testing. It is known that the optimal measurement in such a setting is given by projections onto the positive and negative eigenspaces of the operator,
\begin{equation}
\begin{split}
\hat{H} &= b_2 \dyad{\psi(x_2)} - b_1 \dyad{\psi(x_1)},
\end{split}
\end{equation}
which in the $\{\ket{e_{+}},\ket{e_{-}}\}$ basis assumes the matrix representation,
$$
H = \frac{1}{2}
\begin{bmatrix}
\kappa(1+\varphi) & \sqrt{1-\varphi^2}\\
\sqrt{1-\varphi^2} & \kappa(1-\varphi)
\end{bmatrix}.
$$
It can further be shown that the matrices $H$ and $L_{\kappa}$ are {\em similar matrices} - they share the same eigenvectors. This implies that the YKL measurement and the SLD measurement for the parameter $\kappa$ are identical.

In Fig. \ref{fig: Two-Source Brightness POVM} we show the two-state geometry of the SLD/YKL eigenvectors $\ket{\upsilon_1}$ and $\ket{\upsilon_2}$. These are the optimal projectors for discriminating $\ket{\psi(x_1)}$ and $\ket{\psi(x_2)}$. We also depict their spatial mode representation $\upsilon_1(x) = \braket{x}{\upsilon_1}$ and $\upsilon_2(x)=\braket{x}{\upsilon_2}$ over the image plane. The eigenvectors vary both as a function of the brightness bias $\kappa$ and the emitter half-separation $s=(x_{2}-x_{1})/2$. In particular, the optimal modes privilege the brighter source - they are configured to enhance classification accuracy of the brighter source at the expense of increasing classification error of the dim source. 

\subsection*{Qualitative Analysis of $K>2$ Emitters}

Whether the YKL measurement saturates the full brightness QCRB for $K > 2$ emitters remains an open question. Nonetheless, the structural correspondence established in the two-emitter case motivates the use of YKL-SPADE measurement for brightness estimation 
in larger emitter ensembles.

\begin{figure*}
    \centering
    \includegraphics[width=.9\linewidth]{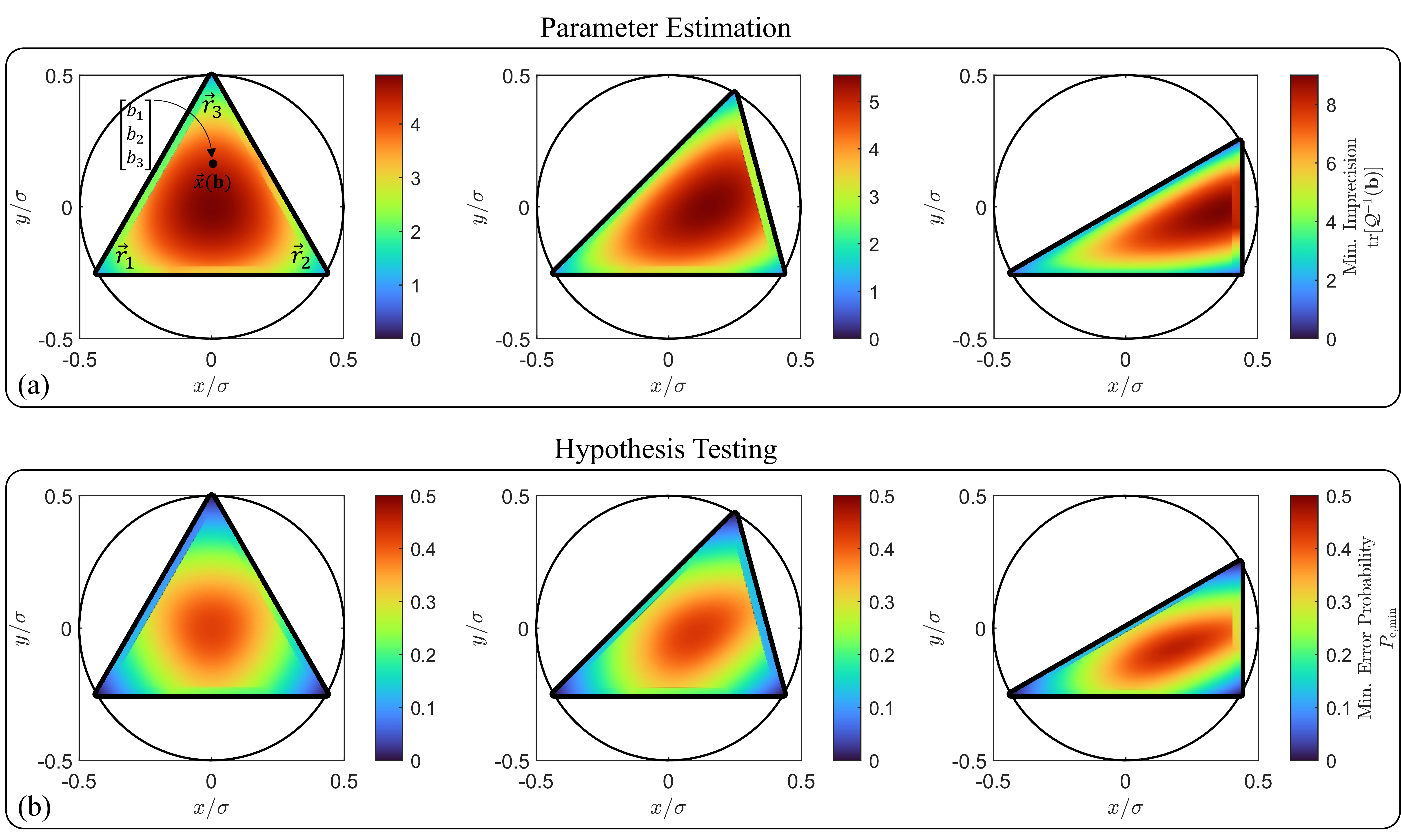}
    \caption{Visual representations of the fundamental performance bounds on \textbf{(a)} brightness estimation imprecision given by the trace of the quantum Cram\'er-Rao bound, and \textbf{(b)} the minimum error probability for quantum hypothesis testing. Each sub-figure depicts a different configuration of three emitters lying within a diffraction-limited spot denoted by the circumscribing circle. The geometry of the emitters $\vec{r}_1,\vec{r}_2,\vec{r}_3$ define the vertices of a triangle. Every point within the triangle interior is associated with a barycentric coordinate $\vec{x}(\mbf{b}) = b_1 \vec{r}_1 +b_2 \vec{r}_2 +b_3 \vec{r}_3$ which forms a bijective map with the brightness parameters $\mbf{b}$. The centroid of each triangle corresponds to system of equally bright emitters, the corners correspond to scenes with one dominant emitter, and the edges correspond to scenes with two active emitters with the third opposing emitter turned off. The minimum imprecision and the minimum error probability metrics vary in much the same way over the interior of the simplex. These metrics indicate that estimating the brightness for systems of equally-bright emitters is more challenging than for systems exhibiting  unequal brightness. The geometry of the emitter configuration also impacts the quantum bounds. If there exists a closely-spaced emitter pair (right-most geometry) the imprecision is largest when the emitters forming this proximal pair are dominant in brightness.}
    \label{fig: Imprecision and Error Prob}
\end{figure*}
To assess how well this heuristic tracks the quantum limit for $K > 2$, we 
compare two complementary figures of merit: the minimum total brightness 
imprecision $\sigma_{b}^{2} = \tr[\mathcal{Q}^{-1}_{bb}(\bm{\theta})]$, which 
sets the QCRB on the sum of brightness estimation variances with positions 
treated as known nuisance parameters, and the minimum probability of 
misclassification $P_{\mathrm{e,min}}$ achieved by the YKL measurement. 
For a three-emitter ensemble, Fig.~\ref{fig: Imprecision and Error Prob} 
depicts both quantities over the full brightness simplex, parameterized 
by the barycentric coordinates $\vec{x}(\mbf{b}) = \sum_k b_k\vec{r}_k$ 
within the triangle formed by the emitter positions. Across the entire 
simplex, the two quantities exhibit strikingly similar qualitative structure: 
both reach their maximum when the emitters are equally bright, and increase 
further when a closely-spaced emitter pair is dominant. This global 
correspondence suggests that minimizing classification error and minimizing 
estimation uncertainty are effectively aligned objectives even when a 
closed-form analytical equivalence is unavailable, supporting the practical 
use of YKL-SPADE beyond the two-emitter case.

\begin{figure*}
    \centering
    \includegraphics[width=\linewidth]{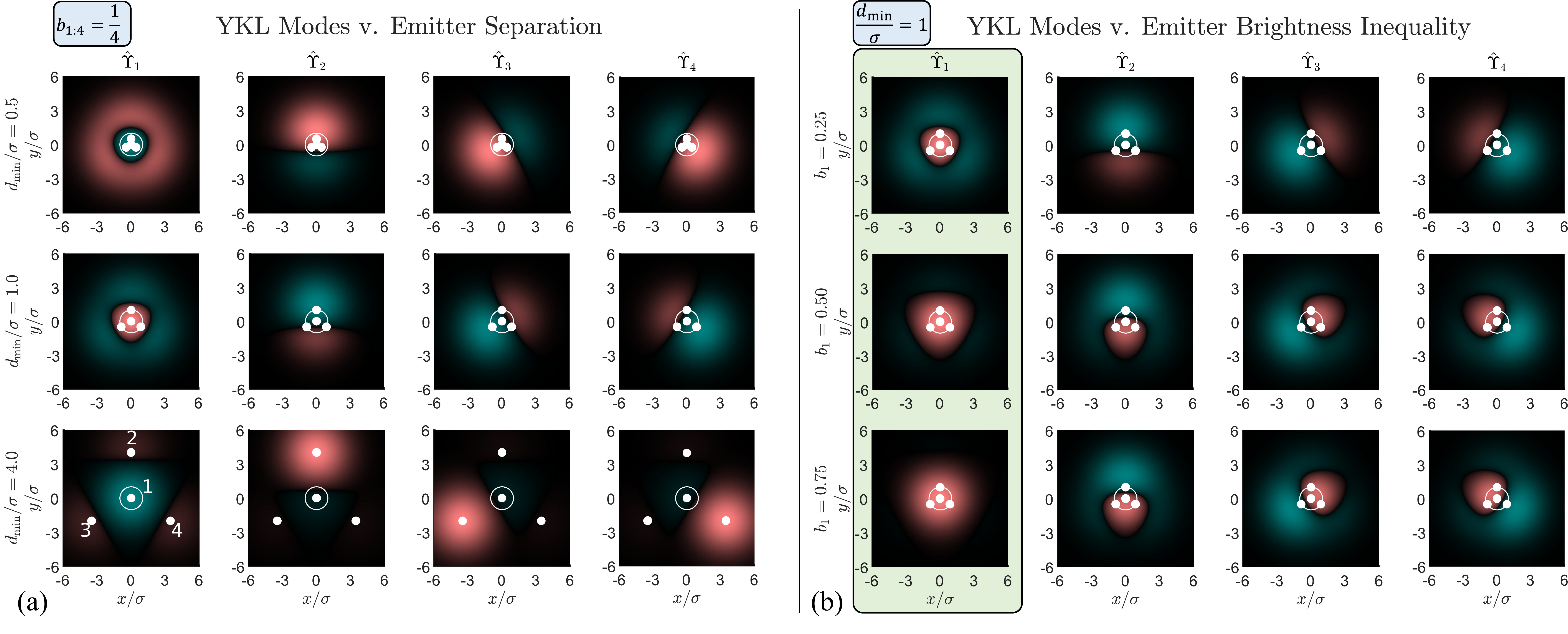}
    \caption{Depiction of the transverse profile of the YKL modes for a constellation consisting of $K=4$ emitters configured as an equilateral triangle with an emitter in the center (emitter positions indicated by white dots). \textbf{(a)} Depicts the YKL modes (for equally bright emitters) as the emitters transition from being closely bunched to well separated. In the limit of large separations, we observe that the YKL modes approximate displaced PSF modes centered at the emitters themselves. Well-separated the states $\ket{\psi(\vec{r}_k)}$ excited by each emitter are nearly orthogonal such that the optimal measurement basis is approximately $\ket{\psi(\vec{r}_k)}$ itself. As the emitter separation enters the sub-diffraction regime (top row) the YKL modes become non-trivial. The mode $\upsilon_k(\vec{x})$ strikes a  balance between coupling as much light as possible from the $k^{\rm th}$ emitter while maintaining orthogonality to the other modes. For this reason, we observe primary lobes of the modes oriented in the direction of the emitter they seek to discriminate. \textbf{(b)} YKL modes for a sub-diffraction emitter ensemble where the brightness of central emitter $k^{\star}=1$ is progressively increased. In the limit where the central becomes dominant, the YKL mode $\nu_1(\vec{x})$ converges to the displaced PSF mode $\psi(\vec{x}-\vec{r}_1)$ indicating that YKL measurement prioritizes correctly discriminating photons emitted by the second source at the expense of erroneously discriminating photons from the other sources.}
    \label{fig: YKL Modes}
\end{figure*}

Further geometric intuition about the YKL measurement is provided by 
Fig.~\ref{fig: YKL Modes}, which depicts the transverse profiles of the YKL 
modes at the image plane for a gaussian PSF as a function of emitter separation and brightness. In the sub-diffraction 
regime, the $k$-th YKL mode $\upsilon_k(\vec{x})$ develops a multi-lobed 
spatial structure whose primary lobe is oriented toward emitter $k$, 
maximizing the coupling $|\braket{\upsilon_k}{\psi(\vec{r}_k)}|^2$ while 
maintaining mutual orthogonality among the modes. As the ensemble separation 
grows beyond the diffraction limit, the YKL modes converge to the individual 
PSF modes $\ket{\psi(\vec{r}_k)}$, consistent with the fact that 
well-separated emitter states become nearly orthogonal and trivially 
discriminable. When one emitter dominates in brightness, its associated YKL 
mode converges to the PSF mode centered at that emitter's location, while 
the remaining modes reorganize to preserve orthogonality. This reflects a 
fundamental trade-off intrinsic to the YKL measurement: it allocates 
discrimination resources preferentially to the brightest emitter at the cost 
of reduced sensitivity to dimmer members of the ensemble.

\section*{Conclusions and Outlook}
\label{sec: Conclusions and Outlook}
We have introduced a protocol for passive super-resolution quantum sensing based on spatial mode sorting. This approach provides a non-invasive alternative for high-resolution imaging in NV-sensing applications and demonstrates significant advantages over direct imaging techniques, particularly in complex environments with multiple vacancy centers.

In our framework, brightness acts as an intermediary parameter through which the sensing protocol infers properties of a target field. We show that well-engineered SPADE measurements, defined by the PAD modes and the YKL modes, extract information more efficiently than direct imaging when estimating emitter positions and brightness within a sub-diffraction ensemble. To demonstrate the utility of this enhanced information efficiency in the context of quantum sensing, we applied our protocol to two practical scenarios: CW-ODMR and Rabi oscillation imaging. In both, we observe substantial improvements in the estimation accuracy of key target parameters - namely, the Zeeman and Rabi frequencies. Deploying our protocol for other sensing applications with well-defined photoluminescence response models is straightforward. We also emphasize that our protocol relies entirely on passive optical pre-processing of the incoming light without any active modulation or illumination of the sample environment. Thus, this method could in principle be combined with other active sensing protocols to compound super-resolution performance. 

Looking forward, several immediate extensions of our protocol appear ripe for investigation. One direction involves adaptively allocating photon resources $N_1^{(\gamma)}$ and $N_2^{(\gamma)}$ based on real-time estimates of the parameter vector $\bm{\theta}_{\gamma}$. Such an adaptive scheduler could dynamically switch between measurements to maximize the accrual of information about the target field $\phi$. Another promising extension would generalize the protocol to wide-field imaging, where the emitters are densely packed. This would require extending our framework from a point emitter cluster model to a continuous emissive distribution as shown in \cite{Tsang:2019_QuantumBoundsIncoherentImaging,Tsang:2020_QuantumSemiparametric,Tsang:2021_QuantLimImaging_ParametricSubmodel}. Fortunately, several works demonstrating the utility of SPADE for imaging extended objects have already shown promise in this direction \cite{Matlin:2022_ExtendedObjectSPADE,Bearne:2021_ConfocalSPADE}. This extension would enable super-resolution sensing with atomic vacancies on various scales.

While the aforementioned extensions may broaden the application space for SPADE-enhanced quantum sensing, various real-world challenges could hamper their utility in experimental settings. Currently, our analysis only accounts for photon shot noise and neglects other noise sources such as detector read noise, dark current, or modal cross-talk which may limit performance in real-world implementations. Bearing these promising directions and important hurdles in mind, we believe that our work lays the foundation for a scalable, non-invasive approach to super-resolution imaging-based atomic sensing.
\newpage

\section*{Acknowledgments}
ND acknowledges support from the US National Science Foundation (NSF) Graduate Research Fellowship under Grant No. DGE-2137419. SG and DD acknowledge funding under the US Air Force Office of Scientific Research (AFOSR) contract number FA9550-22-1-0180 for sponsoring this research. KS acknowledges funding from Department of Science and Technology (DST) Indian National Quantum Mission (NQM) and the AFOSR Asian Office of Aerospace Research and Development (AOARD) grant number FA2386-24-1-4053. AM acknowledges fellowship from I-Hub Chanakya Doctoral Fellowship.

\section*{Author Contributions}
The original concept was conceived by SG and KS and was further refined through discussions with ND, AM, and DD. ND carried out all calculations and simulations in close consultation with the co-authors. DD contributed to the development of the sensing modality, while AM assisted in formulating the initial application of SPADE for brightness estimation. ND drafted the manuscript with input from all authors. SG oversaw the development of the theoretical framework, while KS guided its integration into NV-based magnetometry. Both KS and SG provided overarching supervision and direction throughout the project.

\section*{Data Availability Statement}
Data underlying the results presented in this paper are available in Ref. \cite{Deshler:2025_NV_project_repo}.

\section*{Disclosures}
The authors declare no conflicts of interest.

\bibliography{references}
\bibliographystyle{unsrt}

\newpage
\onecolumngrid
\textbf{Supplementary Materials: Quantum Limited Spatial Resolution of NV-Diamond Magnetometry}

\section{Two Emitters}
\label{SM: Two Emitters}
The intuition underpinning our sensing protocol provided in the main text is largely founded on insights gained from exploring the quantum information limits of sensing with two emitters. In this supplementary material, we:
\begin{enumerate}
    \item Compare the classical Fisher information (CFI) of various measurement strategies against the quantum Fisher information (QFI).
    \item Derive a correspondence between the symmetric logarithmic derivative (SLD) measurement and the YKL measurement for brightness estimation.
    \item Extend our protocol to the Bayesian setting and explore dynamic allocation of photon numbers to different measurement stages.
    \item Introduce derivations that simplify numerical computations in the case of $K>2$ emitter ensembles.
\end{enumerate}

\begin{figure}[h]
 \centering
\includegraphics[width=0.4\linewidth]{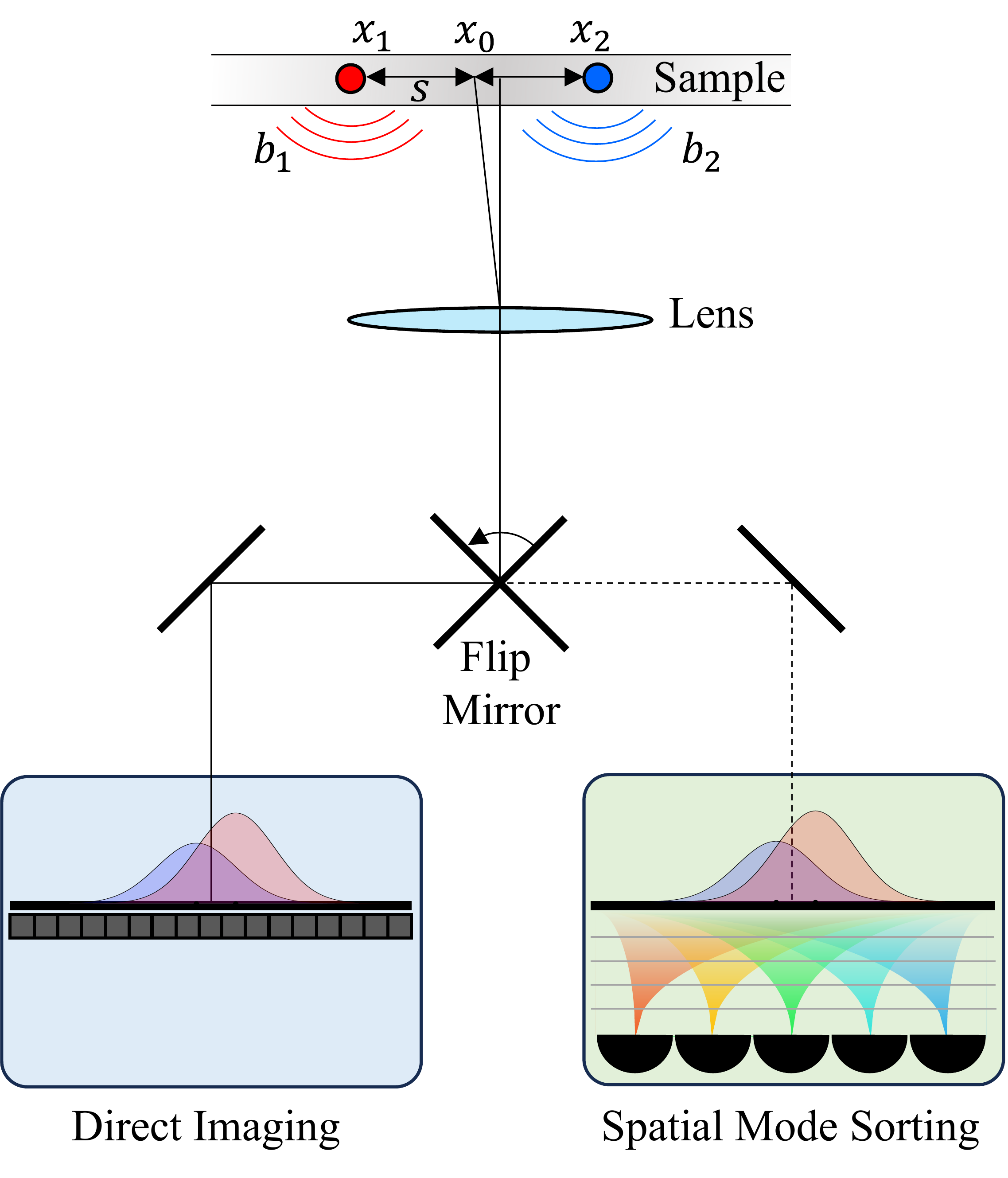}
 \caption{Schematic of two-emitter scene and measurement apparatus for superresolution sensing.}
 \label{fig: Two Source System}
\end{figure}

We consider a simplified system of $K=2$ sub-diffraction emitters separated along one dimension as depicted in Figure \ref{fig: Two Source System}. We assume the imaging system is characterized by a shift-invariant Gaussian point spread function (PSF)  $\psi(x) = (2\pi\sigma^2)^{-\frac{1}{4}}\exp(-\frac{x^2}{4\sigma^2})$. The positions of the emitters are given by $\mbf{r}=[x_1,x_2] \in \mathbb{R}^{2}$ with relative brightness parameters $\mbf{b}=[b_1, b_2] \in S_2$ such that  $b_1 + b_2 = 1$. Under a change of variables, a convenient reparameterization of the two-emitter system as,
\begin{equation}
x_0 = \frac{x_2+x_1}{2}, \quad s = \frac{x_2-x_1}{2}, \quad \kappa = \frac{b_2 - b_1}{2},
\label{eqn: 2-Source Parameters}
\end{equation}
where $x_0\in (-\infty,\infty)$ is the geometric midpoint of the sources, $s \in (0,\infty)$ is the half-separation of the sources, and $\kappa \in (-\frac{1}{2},\frac{1}{2})$ is the bias in the brightness of the sources. Without loss of generality, we assume $x_2 > x_1$ such that $s\in \mathbb{R}_{+}$ is always positive. The brightness bias $\kappa$ is negative if the left source (located at $x_1$) is brighter than the right source and positive if the right source is brighter than the left. For equally bright sources, $\kappa = 0$. The single photon state associated with this system is given by,
\begin{equation}
    \rho(\bm{\theta}) = b_1 \dyad{\psi(x_1)} + b_2 \dyad{\psi(x_2)}.
    \label{eqn: 2-Source State}
\end{equation}
This state has been the object of intense interest within the quantum-inspired super-resolution imaging literature as it beckons toward an applicable quantum multi-parameter estimation task with rich analytical insights. In a broad sense, our goal throughout the following sections will be to explore estimation strategies for the parameters $\bm{\theta} = [x_0,s,\kappa]$.

\subsection{Fisher Information}
\label{SM: Two Sources Fisher Information}
Quantum information theory provides a framework for determining the fundamental limit of precision with which any parameter of a quantum state can be estimated. In particular, for any locally unbiased estimator $\check{\bm{\theta}}$ acting on measurements of the parametric state $\rho(\bm{\theta})$, the covariance matrix of the estimator is subject to the quantum Cram\'er-Rao bound (QCRB):
\begin{equation}
    \text{cov}[\check{\bm{\theta}}] \geq \frac{1}{N}\mathcal{Q}^{-1}(\bm{\theta}),
    \label{eqn: QCRB}
\end{equation}
where $N$ is the number of identical state copies available and $\mathcal{Q}$ is the quantum Fisher information matrix (QFIM) with entries,
\begin{equation}
    \mathcal{Q}_{ij}(\bm{\theta}) = \Tr[\rho(\bm{\theta})(\hat{L}_{\theta_i}\circ \hat{L}_{\theta_j})],
\end{equation}
where $\hat{A}\circ \hat{B} = \frac{1}{2}(\hat{A}\hat{B}+\hat{B}\hat{A})$ is the Jordan product and $\hat{L}_{i}$ are the symmetric logarithmic derivative (SLD) defined implicitly by the Lyapunov equation,
\begin{equation}
\partial \theta_j \rho(\bm{\theta}) = \hat{L}_{\theta_j} \circ \rho(\bm{\theta}),
\end{equation}
where we introduce the short-hand $\partial_{\theta_j} = \partial/\partial_{\theta_j}$. For a given measurement $\hat{\mathcal{M}}_k$ the CFIM is given by the matrix,
\begin{subequations}
    \begin{align}
        \mathcal{I}_{ij}(\bm{\theta}) &= \sum_{k} \frac{\big( \partial_{\theta_i}p_k(\bm{\theta})\big)\big( \partial_{\theta_j}p_k(\bm{\theta})\big)}{p_k(\bm{\theta})}, \\
        p_{k}(\bm{\theta}) &= \Tr[\rho(\bm{\theta})\hat{\mathcal{M}}_{k}].
    \end{align}
\end{subequations}
For the two-emitter system parameterized by $\bm{\theta} = [x_0,s,\kappa]$, the QFIM was found by \cite{Rehacek:2017} to be
\begin{equation}
Q(\bm{\theta}) = \frac{1}{\sigma^2}
    \begin{bmatrix}
        1 - (\frac{s\varphi}{\sigma})^2(1-4\kappa^2) & 2\kappa & 2 s \varphi^2 \\
        2\kappa & 1 & 0\\
        2 s \varphi^2 & 0 & \frac{4\sigma^2 (1-\varphi^2)}{(1-4\kappa^2)}
    \end{bmatrix},
    \label{eqn: QFIM}
\end{equation} 
where the constants (for the case of Gaussian PSF) are given by,
\begin{subequations}
\begin{align}
\varphi &= \braket{\psi(x_1)}{\psi(x_2)} = e^{-\frac{1}{2}(\frac{s}{\sigma})^2},\\
        r^2 &= \mel{\psi(x_1)}{\hat{P}^2_{x}}{\psi(x_1)} =\frac{1}{4\sigma^2},
\end{align}
\label{eqn: 2-Source QFIM Constants}
\end{subequations}
with $\hat{P}_x$ being the momentum operator (i.e., $\hat{P}_x = -i\partial_x$ in the position representation). In general, the optimal measurement requires knowing the values of the parameters \textit{a priori}. Our sequential estimation protocol endeavors to circumvent this predicament by applying measurements that are optimal (or near optimal) for each parameter, one at a time. 

\begin{figure*}
 \centering
\includegraphics[width=.8\linewidth]{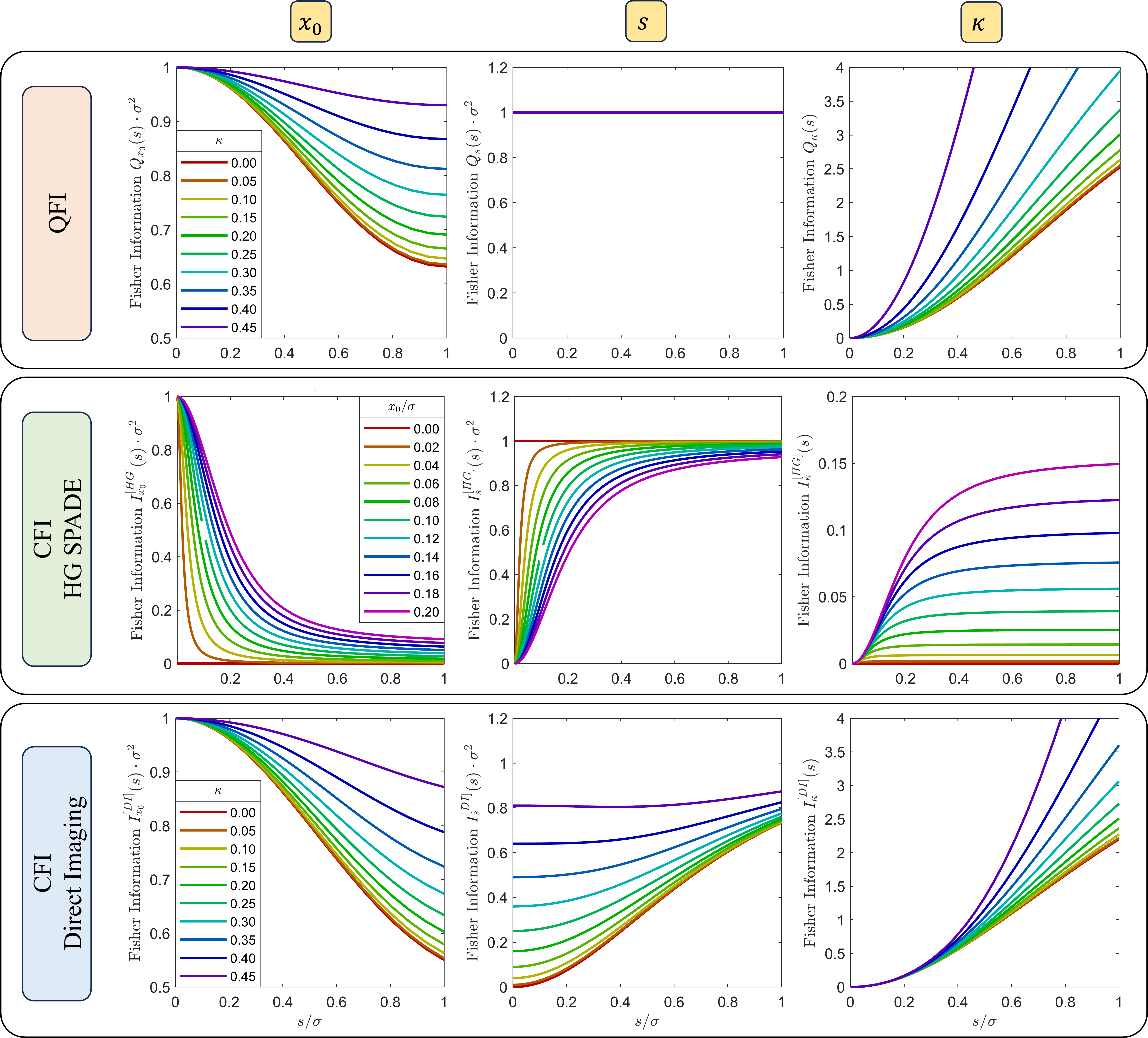}
\caption{Graphs of the quantum and classical fisher information (QFI/CFI) for parameters $(x_0,s,\kappa)$ as a function of the source separation. The CFI is computed for direct imaging and HG SPADE measurements. In particular, these plots depict the diagonal elements of the Fisher information matrices as a function of the source separation $s$ over a sub-diffraction interval. Each colored line in the QFI and Direct Imaging CFI plots correspond to different values of the brightness parameter $\kappa$ with arbitrary value of the midpoint since Fisher information for both of these measures is invariant under a global translation of the coordinate origin (so we need not worry about the value of $x_0$). Meanwhile the different colored lines in the CFI plots of HG-SPADE correspond to different positions of the midpoint relative the optical axis of the mode sorter - for this measurement the Fisher information is not translation invariant. Additionally, for the HG-SPADE subplots we only consider the case of balanced sources $\kappa = 0$.
 The first column shows the Fisher information of the midpoint $x_0$. Examining the QFI, we find that very unbalanced scenes $(|\kappa| \approx .5)$ contain more information about the midpoint that balanced ones $(\kappa \approx 0)$. We further observe that direct imaging nearly saturates the QFI while HG-SPADE significantly under-performs in the regime where the sources are slightly separated. Interestingly, HG-SPADE experiences complete insensitivity to the midpoint in the pathological case where $x_0 = 0$ such that the midpoint and the optical axis of the mode sorter are in perfect alignment.
The second column shows the Fisher information of the separation parameter $s$. Here the QFI is shown to be constant irrespective of all of the parameters. In this case, the Fisher information of HG-SPADE measurements appears to dramatically outperform that of direct imaging in the case where the optical axis of the mode sorter is well-aligned to the midpoint. Meanwhile, the information efficiency of direct imaging improves as the sources become severely imbalanced or when their separation is large. Finally, the third column examines the Fisher information for $\kappa$. As one may expect, the QFI grows with the separation - hence the brightness of well-separated sources may be more accurately reported. Moreover, the QFI for $\kappa$ grows with the imbalance in the brightness itself since the purity of $\rho$ improves. We also find that the classical Fisher information under direct imaging and HG-SPADE are notably sub-optimal compared to the QFI. This motivates the search for another measurement that exhibits better sensitivity to brightness.}
 \label{fig: Fisher Information}
\end{figure*}

In Figure \ref{fig: Fisher Information} we compare the classical fisher information of PAD-SPADE and direct imaging achieved for each of the parameters. Note that DI constitutes a near-optimal measurement for determining the midpoint $x_0$ while PAD-SPADE is an optimal measurement for estimating the half-separation $s$ (assuming ideal pointing) as originally found in \cite{Tsang:2016}'s seminal work. Importantly, neither of these measurements are optimal for estimating the brightness bias $\kappa$. For this reason, we introduce the YKL-SPADE measurement which saturates the QFI bound on $\kappa$ (conditioned on knowledge of $x_0$ and $s$). Later we show that the YKL-SPADE measurement is equal to the SLD measurement for the brightness parameter $\kappa$. This draws an interesting connection between quantum estimation and quantum hypothesis testing. However, unlike DI and PAD-SPADE, which are parameter-independent POVMS, the YKL-SPADE measurement is parameter-dependent, requiring a pre-estimate of $x_0$, $s$ and $\kappa$ in order to configure the SPADE device accordingly.

\subsection{SLD-YKL Correspondence}
\label{SM: Two Sources SLD-YKL Correspondence}
The operational meaning of the mixed state in equation \ref{eqn: 2-Source State} can be described in the following way. Each photon arriving at the image plane is definitively in a pure state, either $\ket{\psi(x_1)}$ or $\ket{\psi(x_2)}$. The supplied state is generated randomly by a classical process with probability $b_1$ and $b_2=1-b_1$ (e.g. a biased coin flip). The mixed state $\rho(\bm{\theta})$ simply describes our classical ignorance about which state was generated. We may therefore imagine a (random) binary sequence of $N$ states which are the true states we receive. For example, one such sequence may look like,
$$
\underbrace{
\ket{\psi(x_1)},\ket{\psi(x_2)},\ket{\psi(x_2)},\ket{\psi(x_1)},\ldots,\ket{\psi(x_1)}}_{N \text{ state sequence}}.
$$
In the limit $N\rightarrow\infty$, the relative frequency with which $\ket{\psi(x_1)}$ ($\ket{\psi(x_2)}$) appears approaches $b_1$ ($b_2$) by Sanov's Theorem. Therefore, brightness estimation appears commensurate to identifying the relative frequency with which each state appeared in the sequence. From this perspective, we may intuit that the optimal measurement for determining the brightness should relate to the optimal hypothesis testing measurement for discriminating $\ket{\psi(x_1)}$ and $\ket{\psi(x_2)}$. Indeed we will prove that the optimal measurement for estimating the brightness bias $\kappa$ given by projections onto the eigenstates of the SLD $\hat{L}_{\kappa}$ is equivalent to the YKL measurement for binary state discrimination. This measurement is also known as the Helstrom measurement in the special case of binary hypothesis testing. 

We begin by calculating the SLD of the brightness parameter $\kappa$ which satisfies the Lyapunov equation, 
\begin{equation}
    \partial_{\kappa} \rho(\bm{\theta})= \hat{L}_{\kappa}\circ \rho(\bm{\theta}).
\end{equation}
Then define the operators 
\begin{equation}
\hat{\Delta}_{\pm} \equiv \dyad{\psi(x_2)} \pm \dyad{\psi(x_1)},
\end{equation}
such that $\rho(\bm{\theta}) = \frac{1}{2}\hat{\Delta}_{+} + \kappa \hat{\Delta}_{-}$ and the Lyapunov equation reduces to,
\begin{equation}
\hat{\Delta}_{-}=\hat{L}_{\kappa}\circ\rho.
\label{eqn: Lyap Brightness Param}
\end{equation}
Defining the orthonormal basis $\{\ket{e_{+}},\ket{e_{-}}\}$:
\begin{subequations}
\begin{align}
\ket{e_{+}} &= \frac{1}{\sqrt{2(1+\varphi)}}(\ket{\psi(x_2)} + \ket{\psi(x_1)}),\\
\ket{e_{-}} &= \frac{1}{\sqrt{2(1-\varphi)}}(\ket{\psi(x_2)} - \ket{\psi(x_1)}),
\end{align}
\end{subequations}
where $\varphi$ is defined in Eqn. \ref{eqn: 2-Source QFIM Constants}. Equivalently, we have
\begin{subequations}
\begin{align}
\ket{\psi(x_1)} &= \sqrt{\frac{1+\varphi}{2}}\ket{e_{+}} - \sqrt{\frac{1-\varphi}{2}}\ket{e_{-}},\\
\ket{\psi(x_2)} &= \sqrt{\frac{1+\varphi}{2}}\ket{e_{+}} + \sqrt{\frac{1-\varphi}{2}}\ket{e_{-}}.
\end{align}
\end{subequations}
Under the ordered representation $\{\ket{e_{+}},\ket{e_{-}}\}$ the operators $\hat{\Delta}_{\pm}$ can be written as matrices,
\begin{equation}
\Delta_{+} = \begin{bmatrix}
    (1+\varphi)  & 0 \\
    0 & (1-\varphi)
\end{bmatrix},\,\, \Delta_{-} = \begin{bmatrix}
    0 & \sqrt{1-\varphi^2}\\
    \sqrt{1-\varphi^2} &0
\end{bmatrix}.
\end{equation}
One can verify that the solution to the Lyapunov equation of \ref{eqn: Lyap Brightness Param} is given by the SLD matrix,
\begin{equation}
L_{\kappa} = \frac{2}{\tau}
\begin{bmatrix}
 -2\kappa(1-\varphi) & \sqrt{1-\varphi^2}\\
 \sqrt{1-\varphi^2} & -2\kappa(1+\varphi)
\end{bmatrix},
\end{equation}
where $\tau \equiv (1-4\kappa^2)$. In the $\{\ket{e_{\pm}}\}$ representation, the SLD has orthogonal eigenvectors, 
\begin{subequations}
\begin{align}
    \bm{\upsilon}_1 &\propto \bigg[ \frac{2\kappa \varphi + \sqrt{1-\tau \varphi^2}}{\sqrt{1-\varphi^2}},1\bigg]^{\intercal}, \\
    \bm{\upsilon}_{2} &\propto \bigg[ \frac{2\kappa \varphi - \sqrt{1-\tau \varphi^2}}{\sqrt{1-\varphi^2}},1\bigg]^{\intercal}, 
    \label{eqn: SLD-Helstrom Projectors}
\end{align}
\end{subequations}
up to normalization factors. The optimal POVM for brightness estimation in the case of two emitters is given by the set of projectors constructed from the eigenstates of the SLD:
\begin{equation}
\{\hat{\Upsilon}_1 = \dyad{\upsilon_1} , \hat{\Upsilon}_2 = \dyad{\upsilon_2}\}.
\end{equation}
This measurement requires knowledge of the brightness bias $\kappa$ as well as the source positions $x_1$ and $x_2$.

With eigenvectors of the SLD in hand, we now turn to the YKL measurement (Helstrom measurement) which admits a closed-form expression in the case of binary state classification. It is known that the optimal measurement is such a setting is given by projections onto the positive and negative eigenspaces of the operator,
\begin{equation}
\begin{split}
\hat{\Delta} &= b_2 \dyad{\psi(x_2)} - b_1 \dyad{\psi(x_1)} \\
&= \frac{1}{2}\hat{\Delta}_{-} + \kappa \hat{\Delta}_{+},
\end{split}
\end{equation}
which in the $\{\ket{e_{+}},\ket{e_{-}}\}$ representation is given by the matrix,
$$
\Delta = \frac{1}{2}
\begin{bmatrix}
2\kappa(1+\varphi) & \sqrt{1-\varphi^2}\\
\sqrt{1-\varphi^2} & 2\kappa(1-\varphi)
\end{bmatrix}.
$$
Note that the Helstrom matrix $\Delta$ and the the SLD matrix $L_{\kappa}$ have the same eigenvectors. Therefore, the YKL measurement and the SLD measurement are identical.

In Fig. \ref{fig: Two-Source Brightness POVM} we show the two-state geometry of the optimal projector states $\ket{\upsilon_1}$ and $\ket{\upsilon_2}$ for discriminating $\ket{\psi(x_1)}$ and $\ket{\psi(x_2)}$. We also depict the spatial modes $\upsilon_1(x) = \braket{x}{\upsilon_1}$ and $\upsilon_2(x)=\braket{x}{\upsilon_2}$ corresponding to these optimal projectors. We observe that the projectors vary both as a function of the brightness bias $\kappa$ and the emitter separation $s$. In particular, the optimal modes privilege the brighter source - they are configured to enhance classification accuracy of the brighter source at the expense of increasing classification error of the dim source.

\subsection{Positional Nuisance Parameters and Calibration vs. Sensing Photon Allocations}
\label{SM: Positional Nusiance Params + Optimal Allocation}
One central feature of our protocol is the use of two different states: $\rho(\bm{\theta}_0)$ with $\bm{\theta_0} = [x_0,s,0]$ in the calibration stage, and $\rho(\bm{\theta})$ with $\bm{\theta} = [x_0,s,\kappa]$ in the sensing stage. Here we seek to explore how the number of calibration stage photons $M$ should scale relative to the number of sensing stage photons $N$. To address this question, we elect to treat the positions of the two emitters as contextual nuisance parameters and optimize the ratio of calibration and sensing stage states to minimize the QCRB of the brightness parameter alone.

To determine the QCRB on the brightness parameter we invoke the additivity of Fisher information for independent state copies and define the total QFIM as,
\begin{equation}
\begin{split}
\bar{\mathcal{Q}}(\bm{\theta}) &= M \tilde{\mathcal{Q}}(\bm{\theta}_0) + N \mathcal{Q}(\bm{\theta})  \\
&= (M+N)\bigg[\beta \tilde{\mathcal{Q}}(\bm{\theta_0})+(1-\beta)\mathcal{Q}(\bm{\theta})\bigg],
\end{split}
\label{eqn: QFIM Calibration + Sensing}
\end{equation}
where $\beta = M/(M+N)$ is fraction of calibration states $\rho(\bm{\theta}_0)$, and
$$
\tilde{\mathcal{Q}}_{ij}(\bm{\theta}_0) = 
\begin{cases}
\mathcal{Q}_{ij}(\bm{\theta}_{0}),& i,j \in \{1,2\} \\
0,  & \text{otherwise}
\end{cases},
$$
is the upper-left $2\times2$ sub-matrix of the QFIM in Eqn. \ref{eqn: QFIM} embedded in a $3\times 3$ matrix by zero-padding the bottom row and right column. We are required to invoke this zero pad because the brightness parameter in the case of the calibration state $\rho(\bm{\theta_0})$ is assumed to be known exactly (balanced emitters).

\begin{figure}
    \centering
    \includegraphics[width=0.5\linewidth]{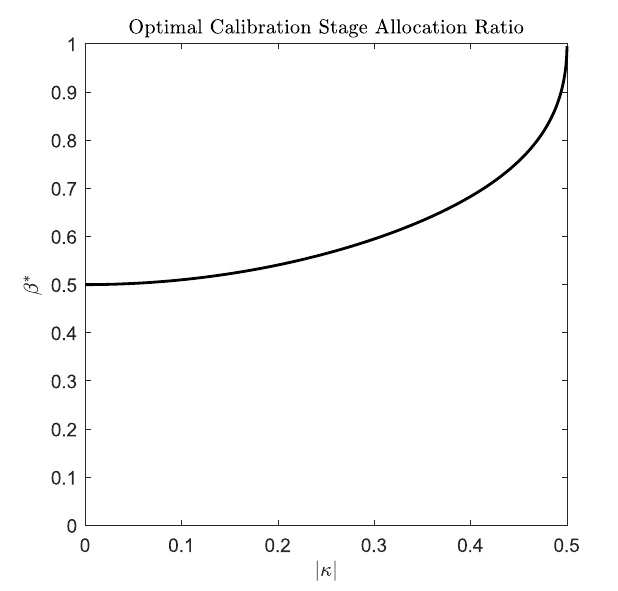}
    \caption{Optimal allocation of resources to the calibration stage versus the sensing stage $\beta = M/(M+N)$ for minimizing the QCRB of brightness estimation in sub-diffraction settings. In the sub-diffraction regime, the optimal $\beta$ is approximately independent of the emitter separation. The curve shown is the solution to Eqn. \ref{eqn: Quadratic Solution for Optimal Calibration-Sensing Allocation} for various values of the brightness $\kappa$. Note that for practical sensing contrasts where $\kappa \leq1/4$ the optimal allocation is approximately an equal balance of calibration and sensing resources $\beta^{*}\approx 1/2$. }
    \label{fig: Optimal Calibration v. Sensing Allocation}
\end{figure}

For a multi-parameter estimation task where the full parameter set can be partitioned into nuisance parameters $\bm{\theta}_{r}$ and target parameters $\bm{\theta}_{b}$ such that $\bm{\theta}=[\bm{\theta}_r,\bm{\theta}_b]$, the QFIM can be written in block form as \cite{Sidhu:2020_Geometric_Parameter_Estimation},
\begin{equation}
\mathcal{Q}(\bm{\theta}) =
\begin{bmatrix}
    \mathcal{Q}_{rr}(\bm{\theta}) & \mathcal{Q}_{rb}(\bm{\theta}) \\
    \mathcal{Q}_{br}(\bm{\theta}) & \mathcal{Q}_{bb}(\bm{\theta})
\end{bmatrix}.
\label{eqn: Block QFIM}
\end{equation}
The Cramer-Rao bound for the target parameters is,
\begin{equation}
\text{cov}[\check{\bm{\theta}}_{b}] \geq [\mathcal{Q}_{bb} - \mathcal{Q}_{br}\mathcal{Q}^{-1}_{rr}\mathcal{Q}_{rb} ]^{-1}(\bm{\theta}).
\label{eqn: QCRB Nuisance-Target Partition}
\end{equation}
Here the off-diagonal blocks of the QFIM encode the sensitivity of the uncertainty target parameters to the uncertainty in the nuisance parameters. Treating the position parameters as nuisance parameters $\bm{\theta}_{r}=[x_0,s]^{\intercal}$ and the brightness parameter as the target $\bm{\theta}_{b} = [\kappa]$, the QCRB on any unbiased brightness estimator found by applying Eqn. \ref{eqn: QCRB Nuisance-Target Partition} to Eqn. \ref{eqn: QFIM Calibration + Sensing} is,
\begin{equation}
\text{var}[\check{\kappa}] \geq
\frac{1}{(N+M)}
\bigg[ 
\frac{(1-c)[1-c\beta^2-h(1-c\beta)]}
{4\beta[(1-c\beta^2)(1-\varphi^2)-h(1-c\beta - (1-\beta)\varphi^2)]}
\bigg],
\label{eqn: Brightness QCRB}
\end{equation}
where $c \equiv 4\kappa^2$ and $h  \equiv (s\varphi/\sigma)^2$.
Note that the inequality \ref{eqn: Brightness QCRB} involves variables (namely $\varphi, \beta, h, c$) that are strictly $\leq 1$ for sub-diffraction scenes. In the sub-diffraction limit $s<<\sigma$, we take $\varphi^2 \rightarrow 1-(s/\sigma)^2$ and $h \rightarrow (s/\sigma)^2$. Here, the QCRB on the variance of the brightness estimator asymptotically approaches,
\begin{equation}
\begin{split}
\text{var}[\check{\kappa}] \gtrsim & \bigg(\frac{1}{N+M}\bigg) \bigg(\frac{\sigma}{s}\bigg)^2 \bigg(\frac{1}{4\beta^2}\bigg) \times \\
&\bigg[\frac{1}{\beta(1-\nu)} - \frac{1}{(1-\nu\beta^2)}\bigg]^{-1}.
\end{split}
\label{eqn: Brightness QCRB Sub-Diffraction Approx}
\end{equation}
From inequality \ref{eqn: Brightness QCRB Sub-Diffraction Approx}, we see that the allocation ratio $\beta$ between the calibration block and the sensing block may be optimized to minimize the QCRB. The optimal allocation ratio involves solving the following depressed quartic equation for $\beta$,
\begin{equation}
    \nu^2 \beta^4  - 2 \nu \beta^2  - 2(1-\nu)\beta + 1 = 0,
    \label{eqn: Quadratic Solution for Optimal Calibration-Sensing Allocation}
\end{equation}
and choosing the solution within the range $[0,1]$. Interestingly, under the sub-diffraction approximation the optimal allocation ratio depends only on the brightness parameter $\kappa$, not on the source separation. Figure \ref{fig: Optimal Calibration v. Sensing Allocation} shows the optimal value of $\beta$ found by solving Eqn. \ref{eqn: Quadratic Solution for Optimal Calibration-Sensing Allocation} for different values of $\kappa$. For moderate contrasts characteristic of real NV sensing experiments (i.e. where the non-radiative transition probability is typically around $\chi = 1/2$ corresponding to a contrast where $|\kappa| \leq 1/4$) we observe that the optimal division of resources between the calibration and sensing stages is approximately $1/2$.

\section{Extensions to Sequential Bayesian Estimation}
\label{SM: Bayesian Protocol}
The measurement protocol and estimation methods we defined in the main text for superresolution vacancy sensing is entirely Fisherian - the parameters are assumed to be deterministic quantities which we seek to estimate. An alternative framework that has found success in non-asymptotic quantum metrology is adaptive Bayesian estimation methods where prior distributions on the parameters capture our immediate uncertainty about the parameters. These priors are iteratively updated with each measurement. Here, we introduce a fully Bayesian formulation of our protocol for the simple case of two emitters which elucidates the primary considerations associated with generalizing our protocol to the Bayesian setting. 

\subsection{Calibration Stage}
\label{SM: Bayesian Protocol - Calibration Stage}
As in the main text, the calibration stage seeks to facilitate source localization. The emitters are assumed to be equally bright $\bm{\theta}_0 = [x_0,s,0]$ such that the state supplied during the calibration stage is,
\begin{equation}
    \rho(\bm{\theta}_0) = \frac{1}{2}\bigg(\dyad{\psi(x_1)} + \dyad{\psi(x_2)} \bigg),
\end{equation}
The probability distribution for the the arrival position of any single photon under the direct imaging measurement is,
\begin{eqnarray}
p(x|x_0,s) &=&\Tr[\rho(\bm{\theta}_0) \hat{\Pi}_x] \nonumber \\
&=& \frac{1}{2}\bigg(|\psi(x-x_1)|^2 + |\psi(x-x_2)|^2 \bigg).
\label{eqn: Direct Imaging PDF}
\end{eqnarray}
Let $\mbf{x}\in\mathbb{R}^{M_1}$ be the collection of independent identically distributed direct imaging measurement outcomes for $M_1$ photons. The joint distribution for this measurement is,
\begin{equation}
p(\mbf{x}|x_0,s) = \prod_{i=1}^{M_1}p(x_i|x_0,s).
\end{equation}
For a sub-diffraction pair of emitters, the maximum likelihood (ML) estimator for the midpoint is well approximated by the arithmetic mean of the photon arrival locations,
\begin{equation}
\check{x}_0(\mbf{x}) \approx \frac{1}{M_1}\sum_{i=1}^{M_1}x_i.
\label{eqn: Midpoint ML Estimator DI}
\end{equation}
Applying the central limit theorem and assuming $s<<\sigma$, the midpoint ML estimator approximates a normally distributed random variable for sufficiently large $M_1$ (see Appendix \ref{apd: Estimator Priors}),

\begin{figure*}
 \centering
 \includegraphics[width=\linewidth]{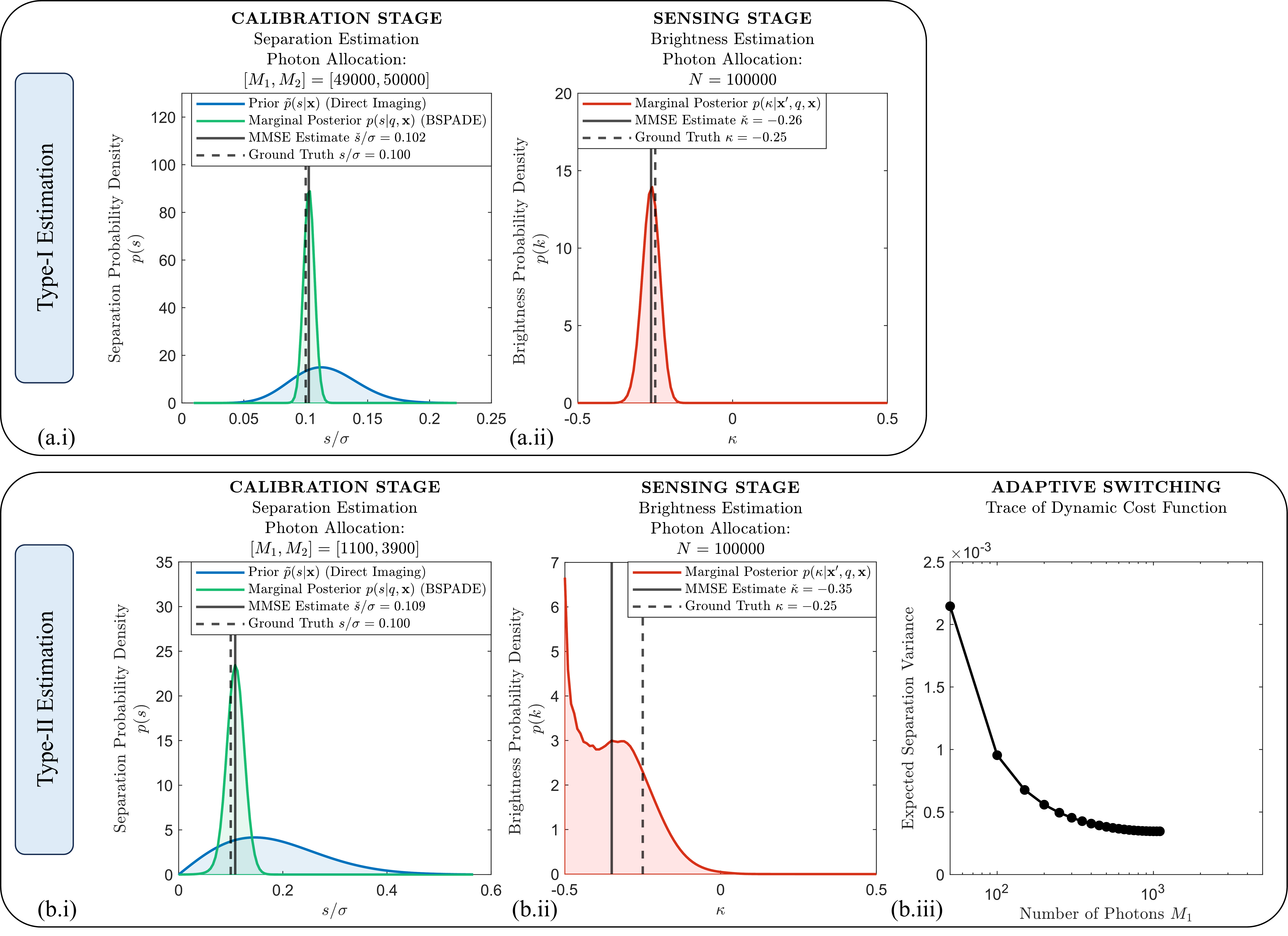}
 \caption{Depiction of posterior updates for the separation parameter $s$ and the brightness parameter $\kappa$ within a Bayesian formulation of our protocol for a sub-diffraction emitter pair. \textbf{(a)} Type-I estimation corresponds to an experimental circumstance whereby the number of photons allowed in calibration is unconstrained. Here, photons are accumulated in the DI and B-SPADE measurements until a desired precision (variance) in the separation posterior is reached. Once this threshold is reached, the protocol transitions to the sensing stage. \textbf{(b)} Type-II estimation places a finite photon budget on the calibration stage. This motivates implementing an adaptive switching rule within the calibration stage, similar to \cite{Grace:2020}, that decides when to transition from direct imaging to B-SPADE. Here the switch occurs if the expected variance in the separation posterior increases after some fraction of the photon budget is reached as shown in \textbf{(b.iii)}. In the limit as the source separation goes to zero, it is known that the optimal switching fraction asymptotically approaches $M_{1}/M = M_{2}/M =1/2$ \cite{Grace:2020}.
 }
 \label{fig: Posterior Updates}
\end{figure*}

\begin{equation}
p(\check{x}_0) = \text{Normal}(\check{x}_0|x_0,\sigma^2/M_1).
\label{eqn: Midpoint MLE DI Distribution}
\end{equation}
The direct imaging measurement also supplies partial information about the separation $s$. In the sub-diffraction regime the ML estimator for the separation may be approximated as (see SM Appendix \ref{apd: Estimator Priors}),
\begin{subequations}
    \begin{align}
    \check{s}(\mbf{x}) &\approx \sigma \sqrt{\check{m}_2 (\mbf{x})-1},
    \label{eqn: Separation ML Estimator DI}\\
    \check{m}_2(\mbf{x})&=\frac{1}{M_1}\sum_{i=1}^{M_1} \bigg(\frac{x_i - \check{x}_0(\mbf{x})}{\sigma}\bigg)^2.
    \end{align}
\end{subequations}
 Note that $\check{m}_2$ is an estimator of the whitened second moment of the distribution $p(x|x_0,s)$ from Eqn. \ref{eqn: Direct Imaging PDF}. Therefore the quantity $\check{m}_2 - 1$ is a measure of how much the variance of $p(x|x_0,s)$ departs from the variance of a Gaussian PDF that would model the direct imaging measurement if the source separation were zero (i.e. if the sources were coincident on each other).\footnote{To avoid non-physical (imaginary) solutions in the approximate estimator of Equation \ref{eqn: Separation ML Estimator DI}, we set $\check{s}=0$ if $\check{m}_2 <1$.} By the method of moments, we derive an approximate probability distribution for the ML estimator of the separation given by (see SM Appendix \ref{apd: Estimator Priors}),
\begin{subequations}
\begin{align}
 p(\check{s}) &= \text{Gamma}\big( (\check{s}/2\sigma^2)|\alpha,\lambda\big) \cdot \frac{\check{s}}{2\sigma^2},\\
\alpha &= \frac{1}{2}(\check{m}_2(\mbf{x})-1)^2\bigg(\frac{M_1^2}{M_1 -1}\bigg),\\
\lambda &= 2(\check{m}_2(\mbf{x})-1)\bigg(\frac{M_1^2}{M_1 -1}\bigg).
\end{align}
\label{eqn: Separation MLE DI Distribution}
\end{subequations}
At this point, we are equipped with a probability distribution on $\check{x}_0$ and on $\check{s}$ given by Eqn. \ref{eqn: Midpoint MLE DI Distribution} and Eqn. \ref{eqn: Separation MLE DI Distribution} respectively. 
We subsequently apply a PAD-SPADE measurement (with $Q=1$ for simplicity) given by the POVM,
\begin{equation}
\{\hat{\Psi}_1 = \dyad{\psi(\check{x}_0)}, \hat{I} -\hat{\Psi}_1\}.
\end{equation}
This measurement, hereafter referred to as Binary SPADE (B-SPADE), projects onto the PSF mode, aligned to the current estimate of the midpoint, and its orthogonal complement. The residual misalignment between the true midpoint and the estimate is a random variable $\varepsilon = x_0-\check{x}_0$ with distribution,
\begin{equation}
    p(\varepsilon) = \text{Normal}(\varepsilon|0,\sigma^2/M_1).
\end{equation}
Upon completing the direct imaging measurement, we treat the parameters $\varepsilon$ and $s$ as random variables so as to leverage the Bayesian estimation framework. All knowledge acquired in the direct imaging stage is captured in the form of priors $\tilde{p}(\cdot)$
\begin{subequations}
\begin{align}
    \tilde{p}(\varepsilon) &= \text{Normal}(\varepsilon| 0, \sigma^2/M_1),
    \label{eqn: Midpoint Prior}\\
    \tilde{p}(s) &= \text{Gamma}((s/2\sigma)^2|\alpha,\beta) \cdot \frac{s}{2\sigma^2}.
    \label{eqn: Separation Prior}
\end{align}
\end{subequations}
Note that priors $\tilde{p}(\varepsilon)$ and $\tilde{p}(s)$ are inherited from the distributions of the ML estimators associated with the direct imaging measurements. The next part of the calibration process invokes the B-SPADE measurement on the remaining $M_2$ state copies. Let $q$ represent the total number of photons detected in the PSF mode. The conditional probability on $q$ is given by a binomial distribution,
\begin{subequations}
    \begin{align}
    p(q|s,\varepsilon) &=\text{Binom}(q|\xi(s,\varepsilon),M_2), \\
    \xi(s,\varepsilon) &= \frac{1}{2}\bigg(e^{-(\frac{\varepsilon + s}{2\sigma})^2} + e^{-(\frac{\varepsilon - s}{2\sigma})^2} \bigg).
    \end{align}
\end{subequations}
After the B-SPADE measurement, we refine our knowledge on the separation by computing the conditional posterior,
\begin{equation}
p(s|\varepsilon) = \frac{p(q|s,\varepsilon)\tilde{p}(s)}{\int p(q|s,\varepsilon)\tilde{p}(s)ds}.
\end{equation}
The integration in the denominator must be done numerically. However, since the binomial distribution is finite dimensional, one need only compute a finite number of integrals.

\subsection{Adaptive Switching in the Calibration Stage}
\label{SM: Bayesian Protocol - Adaptive Switching}
Suppose we begin collecting photons under a direct imaging measurement. At some time $t$ we wish to determine whether it is in our interest to switch to the B-SPADE measurement or not. There are two situations we consider which can be implemented depending on the situational constraints of the calibration stage (e.g. time, system sensitivity, etc.). In the first case, we assume that the experimenter is unconstrained in the number of photons they collect during the calibration stage. That is, they care about getting the brightness estimation right and are willing to spend as much time as needed in the calibration stage to do so, but they also do not want to spend any more time than necessary (type I). The second situation is more constrained - the experimenter has a precise photon budget that they may not go over and the estimation protocol must take this into account when choosing to switch from direct imaging to BSPADE (type II).

\subsubsection*{Type I Constraint}
While continuously collecting photons in the direct imaging stage, we monitor whether the measurements $\mbf{x}^{[t]}$ made up to the current time $t$ satisfy the following requirement for the second moment estimator:
\begin{equation}
\check{m}_2(\mbf{x}^{[t]}) - \zeta \sqrt{\mathbb{V}[\check{m}_2(\mbf{x}^{[t]})]} > 1,
\end{equation}
where $\mathbb{V}[\cdot]$ is the variance of the argument. Given the distribution on the second moment estimator, this is equivalent to the requirement (see SM Appendix \ref{apd: Estimator Priors}),
\begin{equation}
\check{m}_2[\mbf{x}^{[t]}]-\zeta \sqrt{2\frac{M_1[t] -1}{M_1[t]^2}} >1,
\end{equation}
where $M_1[t]$ is the number of photons detected in the direct imaging stage up to time $t$. Here $\zeta$ is a tunable parameter that sets how many standard deviation widths the second moment estimate must be above the second moment of the standard normal. In numerical experiments, we find that $\zeta=2$ gives good results. In general, a larger $\zeta$ ensures that the system stays in direct imaging for more time until sufficient statistics are collected. This heuristic switching criterion roughly ensures that the prior on the separation is well defined since we require $\alpha,\beta>0$. It also implicitly ensures (with high probability) that the pointing error going into the BSPADE stage is smaller than the separation itself. After switching to BSPADE, the experimenter may spend as much time as desired to sharpen the posterior on $s$. 

\subsubsection*{Type 2 Constraint}
In the second constraint case, we have a total number of photons available for the calibration stage $M$. Some amount $M_1$ will be allotted to direct imaging on the fly. After switching, the remaining amount $M_2 = M-M_1$ will be allotted to BSPADE. Our switching criterion from direct imaging to BSPADE will involve the expected cost of switching at the current time step. This is done as follows:

Given the direct imaging measurements collected up to the current time $\mbf{x}^{[t]}$, we construct the priors  $\tilde{p}^{[t]}(\varepsilon)$ and $\tilde{p}^{[t]}(s)$ from Eqns. \ref{eqn: Midpoint Prior} and \ref{eqn: Separation Prior}, respectively. We also take the point estimate of the separation to the mean of the prior $\check{s}^{[t]}= \int \tilde{p}^{[t]}(s) sds$. Our goal will be to compute the expected variance on $s$ assuming we switched to BSPADE at the current time step. To do so, we calculate the conditional posterior,
\begin{equation}
p(s|q,M_2^{[t]},\mbf{x}^{[t]},\varepsilon) =  \frac{p(q|M_2^{[t]},s,\varepsilon )\tilde{p}^{[t]}(s)}{\int p(q|M_2^{[t]},s,\varepsilon )\tilde{p}^{[t]}(s) ds},
\end{equation}
followed by the marginal posterior,
\begin{equation}
p(s|q,M_2^{[t]},\mbf{x}^{[t]}) = \int d\varepsilon p(s|q,M_2^{[t]},\mbf{x}^{[t]},\varepsilon) \tilde{p}^{[t]}(\varepsilon).
\end{equation}

From this we may estimate the variance of $s$ for each possible value of the BSPADE measurement we might expect to encounter,
\begin{equation}
\begin{split}
v^{[t]}_q &= \bigg[\int s^2 p(s|q,M_2^{[t]},\mbf{x}^{[t]})ds\bigg] \\ &-\bigg[ \int s p(s|q,M_2^{[t]},\mbf{x}^{[t]})ds\bigg]^2.
\end{split}
\end{equation}
Finally, as our expected posterior variance on $s$ we take the weighted sum
\begin{equation}
\bar{v}^{[t]} = \sum_{q=0}^{M_2^{[t]}} w^{[t]}_q v^{[t]}_q,
\end{equation}
where the weights are given by the likelihood evaluated at the current point-estimate of the separation:
\begin{equation}
w^{[t]}_q = \int p(q|M_2^{[t]},\check{s}^{[t]},\varepsilon) \tilde{p}^{[t]}(\varepsilon) d\varepsilon.
\end{equation}
It turns out that all of these integrals are relatively tractable to compute numerically due to the low dimensionality of the problem. If we keep track of a growing list of expected variances at each time step $t$, we make the switch as soon as the current expected variance exceeds that of the previous time-step. That is we check if,
\begin{equation}
\bar{v}^{[t]} >\bar{v}^{[t-1]},
\end{equation}
and switch if the inequality holds. Otherwise we continue in direct imaging. The posterior updates performed under type I and type II calibration types are visualized in Fig. \ref{fig: Posterior Updates} for a sub-diffraction emitter pair.

\subsection{Sensing Stage}
\label{SM: Bayesian Protocol - Sensing Stage}
During the sensing stage, we transition back to a direct imaging measurement, armed with a refined knowledge of the separation $s$. The brightness parameter is free to vary with the modulation $\gamma$ such that the parameter vector is $\bm{\theta}_{\gamma} = [x_0,s,\kappa_{\gamma}]$. For simplicity, we will hereafter implicitly assume the dependence $\kappa \leftarrow \kappa_{\gamma}$ such that $\bm{\theta} \leftarrow \bm{\theta}_{\gamma}$ as the procedure is identical for each $\gamma$.
The state supplied during the sensing stage is,
\begin{equation}
\rho(\bm{\theta}_{\gamma}) = \bigg(\frac{1}{2}-\kappa\bigg)\dyad{\psi(x_1)} + \bigg(\frac{1}{2}+\kappa \bigg)\dyad{\psi(x_2)}.
\end{equation}
We introduce a flat prior on the brightness parameter $\kappa$ which preserves its viable domain,
\begin{equation}
    \tilde{p}(\kappa) = \text{Unif}\left(\kappa|-\frac{1}{2},\frac{1}{2}\right), \quad \kappa \in[-1/2,1/2].
\end{equation}
Let $\mbf{x}' \in \mathbb{R}^{N_1}$ denote the collection of direct imaging measurement outcomes observed in the sensing stage. We compute the conditional posterior on the brightness bias as,
\begin{equation}
p(\kappa|s,\varepsilon) = \frac{p(\mbf{x}'|\kappa,s,\varepsilon)\tilde{p}(\kappa)}{\int p(\mbf{x}'|\kappa,s,\varepsilon)\tilde{p}(\kappa) d\kappa}.
\label{eqn: Brightness Posterior}
\end{equation}
In the sub-diffraction regime, the posterior on $\kappa$ can be approximated as (see SM Appendix \ref{apd: Sensing Stage DI Likelihood}),
\begin{equation}
p(\kappa|s,\varepsilon) \approx \frac{\prod_{i=1}^{N_1} [1 + \frac{2\kappa s}{\sigma^2}(x_i'-\varepsilon)]}{\int \prod_{i=1}^{N_1} [1 + \frac{2\kappa s}{\sigma^2}(x_i'-\varepsilon)]\tilde{p}(\kappa)d\kappa},
\end{equation}
which is numerically stable under large $N_1$. We can marginalize the posteriors on $s$ and $\kappa$ to get,
\begin{subequations}
\begin{align}
    p(\kappa) &= \iint p(\kappa|s,\varepsilon)p(s|\varepsilon)p(\varepsilon)dsd\varepsilon,\\
    p(s) &= \int p(s|\varepsilon)p(\epsilon)d\epsilon.
    \label{eqn: separation posterior}
\end{align}
\end{subequations}
With these distributions in hand, we may compute the minimum mean-squared error (MMSE) estimator of the separation $s$ and the brightness parameter $\kappa$:
\begin{subequations}
\begin{align}
 \check{s} &= \int s p(s) ds, \,{\text{and}}\\
 \check{\kappa} &= \int \kappa p(\kappa) d\kappa.
  \label{eqn: MMSE Estimator}
 \end{align}
\end{subequations}

\begin{figure*}
    \centering
    \includegraphics[width=\linewidth]{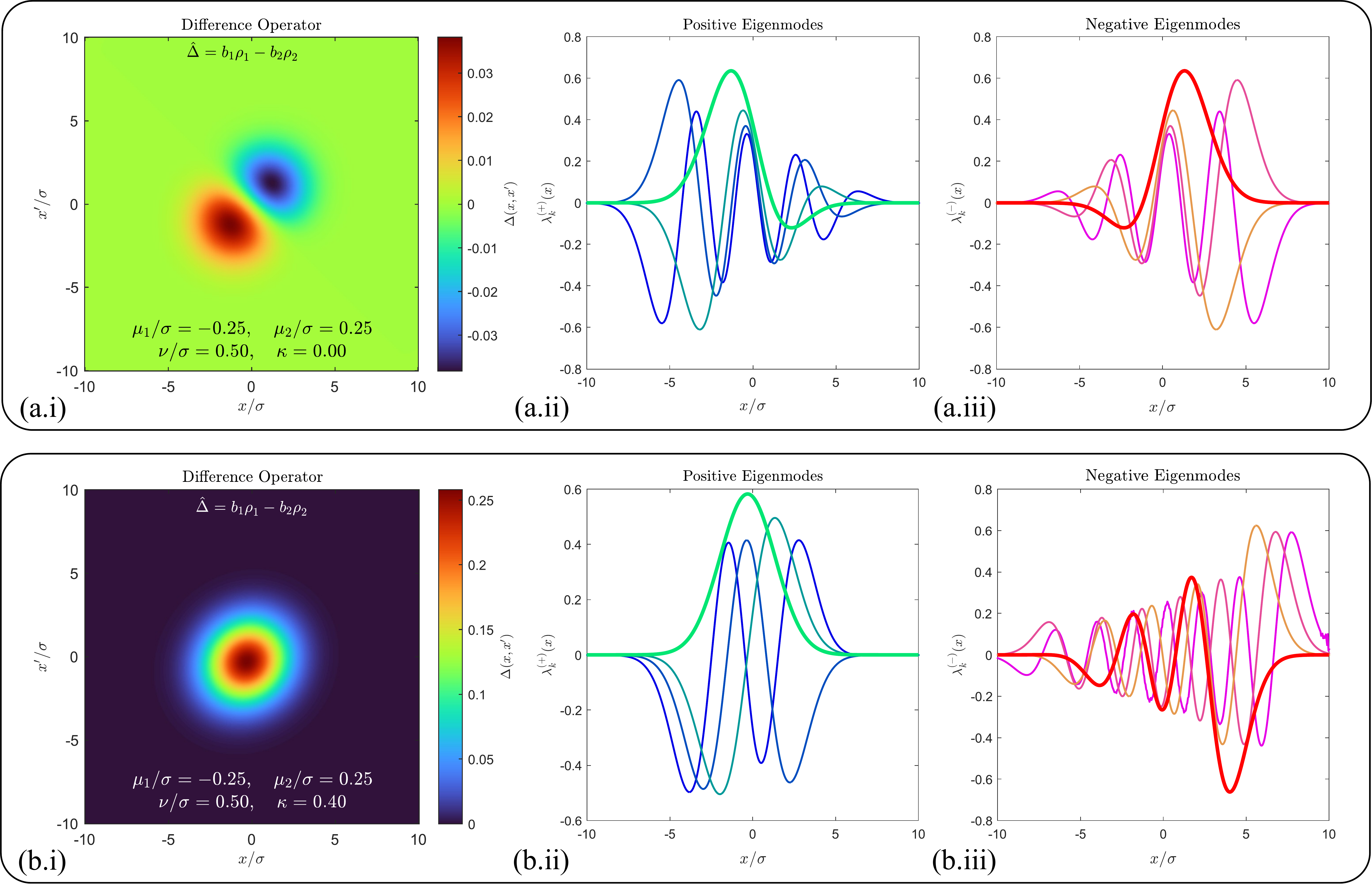}
    \caption{Eigenspectra of the two-source difference operator of Eqn. \ref{eqn: Difference Operator} under \textbf{(a)} balanced source brightness $(\kappa = 0)$ and \textbf{(b)} unbalanced source brightness $(\kappa = 0.4)$. Here the positions $x_1$ and $x_2$ of the point sources are assumed to be classical Gaussian random variables with probability densities given by Eqns. \ref{eqn: Two-Source Gaussian PDFs}. The standard deviation $\nu$ of these distributions characterizes our uncertainty about the emitter positions within a Bayesian estimation framework. Here, the mean emitter positions $\mu_1$ and $\mu_2$ correspond to a sub-diffraction emitter separation in expectation $\mu_2-\mu_1 = \sigma/2$. \textbf{(a/b.i)} Shows the difference operator represented in the position basis. \textbf{(a/b.ii)} and \textbf{(a/b.iii)} show the four most dominant positive and negative eigenmodes (modes with the largest eigenvalue magnitudes) of the difference operator respectively in the image plane. Note that the solid line modes represent the most dominant mode. These are equivalent to the optimal YKL modes derived in Eqn. \ref{eqn: SLD-Helstrom Projectors} and depicted in Fig. \ref{fig: Two-Source Brightness POVM}.}
    \label{fig: Difference Operator Eigenspectra}
\end{figure*}

To define the YKL measurement in the Bayesian context, there are two options: The first option exactly follows the main text and defines a POVM $\{\hat{\Upsilon}_0^{(\gamma)} = \hat{I}-\hat{\Upsilon}_1^{(\gamma)} - \hat{\Upsilon}_2^{(\gamma)}, \hat{\Upsilon}_1^{(\gamma)},\hat{\Upsilon}_2^{(\gamma)}\}$ based on the current parameter estimates. Here,  $\hat{\Upsilon}_1^{(\gamma)}$ and $\hat{\Upsilon}_2^{(\gamma)}$ are the SLD-YKL projectors from Eqn. \ref{eqn: SLD-Helstrom Projectors} under the emitter position estimates $\check{x}_1 = \check{x}_0 - \check{s}$ and $\check{x}_2 = \check{x}_0 + \check{s}$ and brightness bias estimate $\check{s}$. The second option is to incorporate our total present uncertainty about the emitter positions and define:
\begin{subequations}
\begin{align}
\rho_1 &= \int p(x_1) \dyad{\psi(x_1)}dx_1,\\
\rho_2 &= \int p(x_2) \dyad{\psi(x_2)}dx_2,
\end{align}
\end{subequations}
where we treat $x_1 = x_0 - s$ and $x_2 = x_0 + s$ as random variables with probability distributions $p(x_1)$ and $p(x_2)$ inherited from the distributions of $x_0$ and $s$ given by Eqns. \ref{eqn: Midpoint MLE DI Distribution} and \ref{eqn: separation posterior} respectively. Note that this approach is reminiscent in spirit to that of \cite{Zhou:2024_Bayesian2Source} where the authors assume a half-gaussian prior on the separation. We then would invoke the YKL/Helstrom measurement which projects onto the positive and negative eigenspace of the difference operator,
\begin{equation}
\hat{\Delta} =   \check{b}_1 \rho_1 - \check{b}_2 \rho_2,
\label{eqn: Difference Operator}
\end{equation}
where $\check{b}_1 = 1/2 - \check{\kappa}$ and $\check{b}_2 =  1/2 + \check{\kappa}$ with $\check{\kappa}$ being the MMSE estimator of Eqn. \ref{eqn: MMSE Estimator}. Figure \ref{fig: Difference Operator Eigenspectra} shows the difference operator represented in the position basis $\mel{x}{\hat{\Delta}}{x'} = \Delta(x,x')$ under a simple Gaussian probability model for the positions $x_1$ and $x_2$:
\begin{subequations}
    \begin{align}
        p(x_1) &= e^{-\frac{(x_1 - \mu_1)^2}{2\nu^2}}/\sqrt{2\pi \nu^2},\\
        p(x_2) &= e^{-\frac{(x_2 - \mu_2)^2}{2\nu^2}}/\sqrt{2\pi \nu^2}.
    \end{align}
    \label{eqn: Two-Source Gaussian PDFs}
\end{subequations}
We also show the first five most dominant spatial modes within the positive and negative halves of the eigenspectra for $\Delta$. We note that most dominant modes (largest absolute eigenvalues) correspond to the binary YKL measurement shown in Fig. \ref{fig: Two-Source Brightness POVM} under the circumstance where the emitters are deterministically at $x_1 = \mu_1$ and $x_2 = \mu_2$.

Since the domain of $x_1$ and $x_2$ in the formulation of $\rho_1$ and $\rho_2$ is all of $\mathbb{R}$, we do not need to invoke a bucket mode in the YKL measurement to capture the entire Hilbert space $\mathcal{H}$. Suppose the YKL measurement projects onto the positive and negative eigenspace of $\hat{\Delta}$ as defined in Eqn. \ref{eqn: Difference Operator}. For a given value of $\gamma$, we may observe $n_{\gamma}$ photons in the $\hat{\Upsilon}^{(\gamma)}_1$ measurement and $N_2^{(\gamma)}-n_{\gamma}$ in the $\hat{\Upsilon}^{(\gamma)}_2$ measurement. This gives a binomial probability distribution of the form,
\begin{subequations}
\begin{align}
    p(n_\gamma|\bm{\theta}) &= \text{Binom}(n_{\gamma}|q_{\gamma}(\bm{\theta}),N_2^{(\gamma)}), \\
    q_{\gamma}(\bm{\theta}) &= \Tr[\hat{\Upsilon}_{1}^{(\gamma)}\rho(\bm{\theta})],
\end{align}
\end{subequations}
which can then be processed to supply a final estimate of the parameter $\kappa_{\gamma}$.

\section{Many Emitters}
\label{SM: Many Emitters}
In this section, we develop convenient mathematical tools that are widely used throughout our numerical simulations to extend our analysis of the $2$-emitter problem to $K$-emitter ensembles. Specifically, we formulate a numerical procedure for representing emitter state vectors in the eigenbasis of the density operator using only the Gram matrix and the state priors. Additionally, we leverage this formulation to efficiently compute the quantum fisher information matrix and the YKL modes for arbitrary emitter ensembles.

\subsection{Eigenbasis Representation of Arbitrary Convex State Expansions}
\label{SM: State Representation}
Consider a density operator given by the convex combination,
\begin{equation}
\rho = \sum_{i=1}^{K} b_i \dyad{\psi_i} .
\end{equation}
We assume the density operator is full-rank which requires that $b_i>0$ and that $\{\ket{\psi_i}\}$ are linearly independent (though not necessarily orthogonal). The density operator always admits a spectral decomposition of the form,
\begin{equation}
\rho = \sum_{i=1}^{K}\lambda_i \dyad{\lambda_i},
\end{equation}
where $\{\ket{\lambda_i} : \braket{\lambda_i}{\lambda_j} = \delta_{ij}\}$ are the orthonormal eigenvectors of $\rho$ and $\{\lambda_i: \lambda_i >0\}$ are the eigenvalues of $\rho$. Moreover, $\{\ket{\lambda_i}\}$ forms a complete orthonormal basis over the support of $\rho$. Our goal is to express the states $\ket{\psi_i}$ in the eigenbasis $\{\ket{\lambda_i}\}$. 

Assume that projections between states $\{\ket{\psi_i}\}$ are known and collected in the Gram matrix $G$,\footnote{In the special case where $\ket{\psi_i} = \ket{\psi(\vec{r}_i)}$ represent the single-photon states of emitters imaged with Gaussian PSF, the Gram matrix assumes the form, $G_{ij} = \exp{-|\vec{r}_i - \vec{r}_j|^2/8\sigma^2}$.}
\begin{equation}
G_{ij} = \braket{\psi_i}{\psi_j}.
\end{equation}
Since $\{\ket{\psi_i}\}$ are linearly independent, $G$ is positive definite and therefore admits a spectral decomposition of the form $G = UDU^{\dagger}$ where $U$ is a unitary matrix and $D = \text{Diag}(d_1,\ldots,d_K)$ is a diagonal matrix with real-valued non-negative entries.

Next, since $\{\ket{\lambda_i}\}$ forms a complete orthonormal basis over the support of $\rho$, we have the closure relation $\sum_{k=1}^{K} \dyad{\lambda_k}=\hat{I}$ where the identity $\hat{I}$ is on the support of $\rho$. Therefore,
\begin{equation}
\ket{\psi_i} = \hat{I}\ket{\psi_i} = \sum_{k=1}^{K}\braket{\lambda_k}{\psi_i}\ket{\lambda_k}.
\end{equation}
Defining the complex matrix $\Psi\in\mathbb{C}^{K\times K}$ with entries $\Psi_{ki} = \braket{\lambda_k}{\psi_i}$, we may expand the emitter states as,
\begin{equation}
\ket{\psi_i} = \sum_{k=1}^{K}\Psi_{ki}\ket{\lambda_k},
\end{equation}
or, using a vectorized notation, we equivalently have,
\begin{equation}
\begin{bmatrix}
\ket{\psi_1} \\
\vdots \\
\ket{\psi_K} 
\end{bmatrix}
= \Psi^{T} 
\begin{bmatrix}
\ket{\lambda_1} \\
\vdots \\
\ket{\lambda_K}
\end{bmatrix}.
\label{eqn: Matrix Notation for State Expansion}
\end{equation}
Each column of $\Psi$ corresponds to the representation of state $\ket{\psi_i}$ in the eigenbasis of the density operator. Our goal is to determine the (invertible) matrix $\Psi$. The Gram matrix may be written as $G = \Psi^{\dagger}\Psi$. Using the spectral decomposition of the Gram matrix we find that,
\begin{equation}
G = \Psi^{\dagger}\Psi = UDU^{\dagger} =  (UD^{\frac{1}{2}}V^{\dagger})(VD^{\frac{1}{2}}U^{\dagger}),
\label{eqn: Gram Matrix SVD Psi}
\end{equation}
where $V$ is a unitary matrix and we assume the positive roots of $D$ are selected for $D^{\frac{1}{2}}$ such that,
\begin{equation}
 D^{\frac{1}{2}} = \text{Diag}(\sqrt{d_1},\ldots,\sqrt{d_K}),
\end{equation}
as required by the singular value decomposition. In Eqn. \ref{eqn: Gram Matrix SVD Psi}, we can immediately recognize the singular value decomposition of $\Psi$ as,
\begin{equation}
\Psi=VD^{\frac{1}{2}}U^{\dagger}.
\label{eqn: SVD of R}
\end{equation}
Additionally, since $\rho$ is diagonal in the eigenbasis, we have,
\begin{equation}
\Lambda \equiv \text{Diag}(\lambda_1,\ldots,\lambda_) = \Psi B \Psi^{\dagger},
\label{eqn: Density Operator in Eigenbasis}
\end{equation}
where $B = \text{Diag}(b_1,\ldots,b_K)$. At this point, the only remaining degree of freedom in our definition of $\Psi$ the unitary matrix $V$. Inserting equation \ref{eqn: SVD of R} into equation \ref{eqn: Density Operator in Eigenbasis}, we have
\begin{equation}
\Lambda = (VD^{\frac{1}{2}}U^{\dagger})B(UD^{\frac{1}{2}}V^{\dagger})  =   VSV^{\dagger},
\label{eqn: Diagonalization of Density Operator}
\end{equation}
where we have introduced the Hermitian matrix $S=A^{\dagger}A$ with $A = D^{\frac{1}{2}}U B^{\frac{1}{2}}$. From equation $\ref{eqn: Diagonalization of Density Operator}$ we see that the spectral decomposition of $S$ must be $S = V^{\dagger}\Lambda V$.

\subsubsection*{Steps for Representing the Emitter States in the Eigenbasis of the Density Operator}

To summarize, starting from the (non-orthogonal, but linearly independent) states $\{\ket{\psi_i}\}$, their inner products collected in $G$, and the probability weights $P$, we can determine the representation of the states $\ket{\psi_i}$ in the eigenbasis of the density operator by the following procedure:\\

\begin{enumerate}
    \item Compute spectral decomposition of the Gram Matrix $G = UDU^{\dagger}$.
    
    \item Evaluate the matrix $S = A^{\dagger}A$ where $A = D^{\frac{1}{2}}U^{\dagger}B^{\frac{1}{2}}$ 
    \item Compute the spectral decomposition of $S = V^{\dagger}\Lambda V$.
    \item Compute the matrix $\Psi = VD^{\frac{1}{2}}U^{\dagger}$ which is the representation of the states $\ket{\psi_i}$ in the eigenbasis $\ket{\lambda_i}$.
\end{enumerate}


\subsection{Computing YKL States}
\label{SM: Computing YKL States}
Computing the YKL measurement for arbitrary emitter constellations is known to be solvable by semi-definite programming \cite{YKL:1975,Krovi:2015_SymmetricQuantumStateDiscrimination,Bae:2015_QuantumStateDiscriminationApplications}. Here we outline how to compute the YKL state vectors using manifold optimization techniques instead which simplify implementation given the preponderance of efficient manifold optimization packages.

We begin by representing the states $\ket{\psi(\vec{r}_k)}$ as column vectors $\bm{\psi}_k\in \mathbb{C}^{K}$ in the eigenbasis of $\rho(\bm{\theta})  = \sum_{k=1}^{K}b_k \dyad{\psi(\vec{r}_k)}$ via the procedure described in Sec. \ref{SM: State Representation}. Collecting these vectors into a matrix $\Psi$ and defining the diagonal matrix $B = \text{diag}(b_1,\ldots,b_{K})$, we compute the collection of YKL state vectors via manifold optimization on the unitary Lie group $\mathcal{U}(K) = \{U\in\mathbb{C}^{K\times K}: U^{\dagger}U = UU^{\dagger}=I \}$. For an arbitrary projective measurement represented by the matrix $U$, the probability of error (cost function) can be compactly written as,
\begin{equation}
    P_{\text{e}}(U) = 1 - \tr[B\odot|\Upsilon^{\dagger}\Psi|^2],
\end{equation}
where $\odot$ is the Hadamard product and $|\cdot|^2$ is understood to be the element-wise squared modulus of each matrix entry. Minimization of the cost over the Lie group gives the YKL measurement,
\begin{equation}
\Upsilon =\argmin_{U \in \mathcal{U}} P_{\text{e}}(U).
\end{equation}
The solution $\Upsilon \in \mathcal{U}$ is a unitary matrix where each column is a particular YKL state represented in the eigenbasis of $\rho(\bm{\theta})$.
We perform this manifold optimization using the \texttt{manopt} MATLAB package \cite{manopt}. With the YKL measurement matrix in hand, the minimum error probability is $P_{\text{e,min}} = P_{\text{e}}(\Upsilon)$.

\subsection{Quantum Imprecision Limit for Brightness Estimation}
\label{SM: Quant Imprecision Limit for Brightness Estimation}
In addition to the minimum error probability for state discrimination, we may also evaluate the quantum imprecision limit for the collection of brightness parameters in an arbitrary emitter where the emitter position $\mbf{r}$ are assumed to be known nuisance parameters. The minimum imprecision in the brightness parameters is given by,
\begin{equation}
\sigma_{b}^2 = \tr[\mathcal{Q}^{-1}_{bb}(\bm{\theta})],
\label{eqn: Brightness Imprecision}
\end{equation}
which demands computing the QFIM block $\mathcal{Q}_{bb}(\bm{\theta})$ of the brightness parameters defined in Eqn. \ref{eqn: Block QFIM}. To do so, we begin with a matrix definition of the density operator computed as in Sec. \ref{SM: State Representation}:
\begin{equation}
\rho(\bm{\theta}) = \Psi B \Psi^{\dagger}.
\end{equation}
For each brightness parameter, we numerically solve the Lyapunov equation for their SLDs,
\begin{equation}
\frac{\partial}{\partial {b_k}}\rho(\bm{\theta}) = \bm{\psi}_k\bm{\psi}_k^{\dagger} = L_{b_k} \circ \rho(\bm{\theta}).
\end{equation}
The QFIM entries are then found by evaluating,
\begin{equation}
[\mathcal{Q}_{bb}(\bm{\theta})]_{ij} = \tr[\rho(\bm{\theta})(L_{b_i} \circ L_{b_j})].
\end{equation}
Importantly, because $\mbf{b}\in\mathcal{S}_K$ is constrained to live on the probability simplex, there are only $K-1$ independent degrees of freedom. As a result, naively invoking the quantum Cram\'er-Rao bound for an unbiased brightness estimator $\V(\check{\mbf{b}}) \geq \mathcal{Q}_{bb}^{-1}(\bm{\theta})$ has limited utility because it ignores the impact of the constraint on the parameters and the estimator $\check{\bm{\theta}}$ which we expect to be a member of the probability simplex. One may be tempted to force the brightness parameters to obey the constraint equation delegating a single $b_{j} = 1-\sum_{k \neq j}b_{k}$. However, this choice inherently introduces a biased treatment towards certain brightness parameters.

This setting was addressed by Ben-Haim and Eldar \cite{Ben-Haim:2009} who generalized the Cramer-Rao bound to estimation problems where the parameters are subject to a system of constraint equations and the Fisher information matrix is generally positive semi-definite. Applied to our setting, their revised bound is,
\begin{equation}
\V[\check{\mbf{b}}] \geq U(U^{\intercal}\mathcal{Q}_{bb}(\bm{\theta})U)^{-1}U^{\intercal} = ( P\mathcal{Q}_{bb}(\bm{\theta})P)^{+},
\label{eqn: constrained brightness QCRB}
\end{equation}
for any unbiased brightness estimator $\check{\mbf{b}}\in\mathcal{S}_{K}$. Here $A^{+}$ denotes the Moore-Penrose pseudo-inverse of the matrix $A$ and $U\in \mathbb{R}^{K\times(K-1)}$ is an orthogonal matrix $U^{\intercal}U = I_{K-1}$ whose columns form a basis for the tangent space of the probability simplex $\mathcal{T}_{\mathcal{S}_{K}} = \{\mbf{u}\in \mathbb{R}^{K}:\mbf{1}^{\intercal}\mbf{u} = 0\} $. Additionally $P = UU^{\intercal} = I_{K} - \frac{1}{K} \mbf{1}\mbf{1}^{\intercal}$ is the projector onto the tangent space $\mathcal{T}_{\mathcal{S}_{K}}$ and is invariant under the choice of $U$. Hence, while a specific choice of $U$ introduces a particular $(K-1)$-dimensional parameterization of the brightnesses through $\bm{\kappa} = [\kappa_{1},\ldots,\kappa_{K-1}]^{\intercal}$ via,
$$
\mbf{b} = \frac{1}{K}(\mbf{1} + U\bm{\kappa}),
$$
the constrained quantum Cramer-Rao bound of Eq. \ref{eqn: constrained brightness QCRB} is invariant under the choice of $U$.

\subsection{Towards Optimal Measurements for Brightness Estimation}
\label{SM: Optimal Measurements for Brightness Estimation}
In this section, we seek to shed light on the working principles underpinning the efficacy of our protocol for many-emitter sub-diffraction sensing applications. In particular, we draw attention to an interesting connection between quantum state discrimination and quantum parameter estimation in the context of brightness estimation. Additionally, we investigate the transverse structure of the YKL modes as a function of the emitter geometry and their brightness to garner further insight.

In Sec. \ref{SM: Two Sources SLD-YKL Correspondence} we proved that QFI-attaining measurement for brightness estimation of two emitters is equal to the measurement which minimizes the probability of error of discriminating the states supplied by each emitter (i.e. the YKL measurement). Inspired by this correspondence, it is natural to wonder whether: 
\begin{enumerate}
    \item The QFIM for the brightness parameters is achievable in the general case of $K>2$ emitters.
    \item The YKL measurement is an optimal measurement for brightness estimation of $K>2$ emitters.
\end{enumerate}

While these two lines of inquiry presently remain open problems, in Fig. \ref{fig: Imprecision and Error Prob} we provide a visual representation of the correspondence between the QFIM bound and the minimum error probability for the case of three emitters. Specifically, we compute the minimum brightness imprecision $\sigma_b^2(\bm{\theta})$ given in Eqn. \ref{eqn: Brightness Imprecision} and the minimum probability of error $P_{\text{e,min}}$ achieved with the YKL measurement acting on a state $\rho(\bm{\theta})$ comprised of $K=3$ emitters with locations $\vec{r}_1, \vec{r}_2$ and $\vec{r}_3$. The locations of the emitters define the vertices of a triangle. Each point residing in the interior of the triangle can be written as a barycentric coordinate $\vec{x}(\mbf{b}) = b_1 \vec{r}_1 + b_2 \vec{r}_2 + b_3 \vec{r}_3$ which maps bijectively to the brightness values $\mbf{b} = [b_1,b_2,b_3]^{\intercal}$. Points residing closer to the center of the triangle thus correspond to scenes where the emitters are equally bright, while points residing near the vertices of the triangle correspond to scenes where one of the emitters is dominant. This model allows us to visualize the fundamental quantum information limits on brightness estimation over the entire parameter space of a three-emitter scene. We find that the imprecision for estimating brightness is largest when all the emitters are approximately the same brightness. Moreover, the quantum bounds on the estimator precision appear to be dependent on the emitter geometry. If any pair of emitters are closely spaced, then the imprecision bound increases when these two proximal emitters dominate in brightness.

Figure \ref{fig: YKL Modes} illustrates how the transverse profile of the YKL modes varies depending on the emitter geometry and brightness parameters. For sub-diffraction ensembles, the YKL modes exhibit multi-lobed structures that appear to maximize the overlaps $\braket{\upsilon_k}{\psi(\vec{r}_k)}$ while respecting orthogonality $\braket{\upsilon_i}{\upsilon_j}=\delta_{ij}$. As the emitter separation increases beyond the diffraction limit, the YKL modes converge to the shifted gaussian modes $\ket{\upsilon_{k}}\rightarrow \psi(\vec{r}_k)$. In contrast, when we consider varying the brightness of the emitters for a sub-diffraction ensemble, we find that only the YKL mode associated with the brightest source (which we identify as $k^{\star}$) converges as $\ket{\upsilon_{k^{\star}}}\rightarrow \ket{\psi(\vec{r}_{k^{\star}})}$. The remaining modes accommodate themselves to preserve orthogonality. In this way, the YKL measurement privileges coupling to the dominant emitter at the expense of erroneously discriminating photons from less bright emitters.

\appendix

\section{Approximate Midpoint and Separation Priors from Direct Imaging}
\label{apd: Estimator Priors}

The probability distribution for a photon arrival under a direct imaging measurement is given by,
\begin{align*}
p(x|x_0,s) &= \frac{1}{2}\bigg[ \psi((x-x_0)+s))^2 + \psi((x-x_0)-s))^2\bigg] \\
           &= \frac{1}{2\sqrt{2\pi\sigma^2}} \bigg[e^{-\frac{((x-x_0)+s)^2}{2\sigma^2}} + e^{-\frac{((x-x_0)-s)^2}{2\sigma^2}} \bigg]\\
           &= \frac{1}{\sqrt{2\pi\sigma^2}} \bigg[e^{-\frac{(x-x_0)^2}{2\sigma^2}} e^{-\frac{s^2}{2\sigma^2}} \cosh\bigg(\frac{x-x_0}{\sigma^2}\,s\bigg)\bigg].
\end{align*} 
The log-likelihood of the direct imaging measurement $\mbf{x}$ in the calibration stage is,
\begin{equation}
\begin{split}
\mathcal{L}(x_0,s|\mbf{x}) &= M_1\bigg[ -\frac{1}{2}\ln(2\pi\sigma^2)-\frac{s^2}{2\sigma^2}\bigg]\\
&+ \sum_{i=1}^{M_1} \frac{(x_i-x_0)^2}{2\sigma^2} \\ &+ \ln\bigg[\cosh\bigg(\frac{x_i-x_0}{\sigma^2}\, s\bigg)\bigg].
\end{split}
\label{eqn: Log-Likelihood DI}
\end{equation}
In the sub-diffraction regime, we make the approximation ($s \ll \sigma$) such that the first term and logarithmic terms in equation \ref{eqn: Log-Likelihood DI} can be dropped. For the maximum likelihood (ML) estimator of the midpoint, we differentiate the approximate log likelihood with respect to $x_0$ and set the result equal to zero,
$$
0 = \partial_{x_0} \mathcal{L} \approx \frac{1}{\sigma^2}\sum_{i=1}^{M_1}(x_i-x_0).
$$
The ML estimator then simply evaluates to the empirical mean of the photon arrival positions:
$$
\check{x}_0 = \frac{1}{M_1}\sum_{i=1}^{M_1} x_i.
$$
Applying the central limit theorem (assuming sufficiently large $M_1$), the midpoint ML estimator $\check{x}_0$ is a gaussian random variable of the form,
$$
\check{x}_0 \sim \mathcal{N}\left(x_0, \frac{\sigma^2 + s^2}{M_1}\right),
$$
since $\mathbb{E}[x_i] = x_0$ and $\mathbb{V}[x_i] = \sigma^2 + s^2$ for all direct imaging photon arrivals.

For the ML estimator of the separation we differentiate the log-likelihood  with respect to $s$ and set it to zero
$$
0 = \partial_{s}\mathcal{L} = -M_1 \frac{s}{\sigma^2} + \sum_{i=1}^{M_1} \tanh\bigg(\frac{x_i-x_0}{\sigma^2}\,s\bigg) \times \frac{(x_i-x_0)}{\sigma^2},
$$
leading to a transcendental equation,
$$
s = \frac{1}{M_1} \sum_{i=1}^{M_1} \tanh\bigg(\frac{x_i-x_0}{\sigma^2}\,s\bigg) \times (x_i-x_0).
$$
Next we make the sub-diffraction approximation ($s \ll \sigma$) and Taylor expand the $\tanh(\cdot)$ term,
$$
\tanh\bigg(\frac{x_i-x_0}{\sigma^2}\,s\bigg) \approx \bigg(\frac{x_i-x_0}{\sigma^2}\bigg)s - \frac{1}{3}\bigg(\frac{x_i-x_0}{\sigma^2}\bigg)^3 s^3 + \mathcal{O}(s^5),
$$
so that the transcendental equation for the separation ML estimator becomes,
$$
s\approx \frac{1}{M_1} \sum_{i=1}^{M_1} \bigg( \frac{x_i-x_0}{\sigma}\bigg)^2 s - \frac{1}{3\sigma^2} \bigg( \frac{x_i-x_0}{\sigma}\bigg)^4 s^3,  
$$
leading to the approximate ML estimator
$$
\check{s} \approx
\sigma\sqrt{3} \sqrt{ \frac{ \big( \frac{1}{M_1}\sum_{i=1}^{M_1} w_i^2\big) -1 }{(\frac{1}{M_1}\sum_{i=1}^{M_1} w_i^4)}} \quad  \text{where} \quad w_i = \frac{x_i-\check{x}_0}{\sigma}.
$$
In this form, the estimator for the separation assumes an interesting interpretation. Recognizing the $w_i$'s as whitened samples, the numerator inside the square root represents the departure of the empirical variance from the variance of a standard normal distribution $\mathcal{N}(0,1)$. That is when the empirical variance trends larger than $1$, the samples are likely coming from distribution that is wider than a Gaussian, suggesting a separation. In general, we may define the moment estimator for a whitened random variable as,
$$
\check{m}_k = \frac{1}{M_1} \sum_{i=1}^{M_1}w_i^{k},
$$
where $n$ is the number of i.i.d samples of the random variable observed.
Therefore, we have that,
$$
\check{s} \approx \sigma \sqrt{3}\sqrt{\frac{\check{m}_2-1}{\check{m}_4}}.
$$
To simplify the separation estimator even further, we look to evaluate the true moments of the whitened samples. In the sub-diffraction regime, the whitened samples have distribution,
\begin{align*}
p(w_i|x_0,s) &= \frac{1}{\sqrt{2\pi}}e^{-\frac{w_i^2}{2}}e^{-\frac{s^2}{2\sigma^2}}\cosh(\frac{s}{\sigma}w_i)\\
&\approx e^{-\frac{s^2}{2\sigma^2}} \cdot \frac{1}{\sqrt{2\pi}}e^{-\frac{w_i^2}{2}} [1+\frac{s^2}{2\sigma^2}w_i^2].
\end{align*}
The moments $m_k$ of $w_i$ are  the approximated by,
$$
m_{k}  \approx e^{-\frac{s^2}{2\sigma^2}}\bigg[ \mathbb{E}_{\mathcal{N}}[w_i^{k}]+ \frac{s^2}{2\sigma^2}\mathbb{E}_{\mathcal{N}}[w_i^{k+2}]\bigg],
$$
where $\mathbb{E}_{\mathcal{N}}[w_i^{n}] = \frac{n!}{2^{n/2}(n/2)!}$ (for even $n$) are the moments of the standard normal distribution $\mathcal{N}(0,1)$. The two moments of interest for the direct imaging separation ML estimator are
\begin{align*}
    m_2 &\approx  e^{-\frac{s^2}{2\sigma^2}}\bigg[\mathbb{E}_{\mathcal{N}}[w_i^{2}]  + \frac{s^2}{2\sigma^2} \mathbb{E}_{\mathcal{N}}[w_i^{4}] \bigg] = e^{-\frac{s^2}{2\sigma^2}}\bigg[1 + 3 \frac{s^2}{2\sigma^2} \bigg], \\
    m_4 &\approx  e^{-\frac{s^2}{2\sigma^2}}\bigg[\mathbb{E}_{\mathcal{N}}[w_i^{4}]  + \frac{s^2}{2\sigma^2} \mathbb{E}_{\mathcal{N}}[w_i^{6}] \bigg] = e^{-\frac{s^2}{2\sigma^2}}\bigg[3 + 15 \frac{s^2}{2\sigma^2} \bigg]. 
\end{align*}

For sufficiently large $M_1$ we may apply the law of large numbers, taking $\check{m}_4 \approx m_4$. Furthermore, in the sub-diffraction regime, $(s/\sigma)^2<<1$ such that $\check{m}_4 \approx 3$. Then the maximum likelihood estimator for the separation given the direct imaging measurement assumes the simpler form,
$$
\check{s} \approx \sigma \sqrt{\check{m}_{2}(\mbf{x}) - 1}.
$$
Our next objective is to derive a reasonable distribution for $\check{s}$ as we did for $\check{x}_0$. Observe that,
$$
M_1\check{m}_2 = \sum_{k=1}^{M_1} w_i^2  
$$
is the sum of squared of i.i.d random variables which are approximately standard normals - such a sum follows a $\chi^2$ distribution. Thus,
$$
p(\check{m}_2) \approx \frac{1}{M_1} \text{ChiSquared}(M_1 \check{m}_2|M_1),
$$
where the probability density function of the chi-squared distribution is explicitly given by,
$$
f(x) = \text{ChiSquared}(x|k) = \frac{e^{-x/2}}{\Gamma(k/2)2^{\frac{k}{2}}}x^{\frac{k}{2} -1},\,\, x\in[0,\infty).
$$
To derive a distribution on the ML estimator $\check{s}$ of the direct imaging measurement, we will relax the $\chi^2$ distribution and consider the more general family of 
Gamma distributions given by:

$$
f(x) = \text{Gamma}(x|\alpha,\lambda) = \frac{\lambda^{\alpha}}{\Gamma(\alpha)\lambda^{\alpha}}x^{\alpha-1} e^{-\lambda x},
$$
where the mean of the gamma distribution is $\alpha/\lambda$ and the variance is $\alpha/\lambda^2$. This distribution is also conjugate to the Poisson distribution associated with HG mode sorting in the limit of infinite modes. Define series of estimators,
\begin{align*}
\check{\mu}(\mbf{x}) &= \bigg(\frac{\check{s}(\mbf{x})}{2\sigma}\bigg)^2, \\
\check{s}(\mbf{x}) &= \sigma \sqrt{\check{m}_2(\mbf{x}) - 1}, \\
\check{m}_2(\mbf{x}) &= \frac{1}{M_1}\sum_{i=1}^{M_1} \bigg(\frac{x_i - \check{x}_0(\mbf{x})}{\sigma}\bigg)^2, \, {\text{and}}\\
\check{x}_0(\mbf{x}) &= \frac{1}{M_1} \sum_{i=1}^{M_1} x_i,
\end{align*}
such that $\check{\mu}(\mbf{x}) = \frac{1}{4}(\check{m}_2(\mbf{x})-1)$. We will apply the gamma prior to $\check{\mu}(\mbf{x})$ as this estimator is the rate parameter for the Poisson distribution associated with HG mode sorting assuming perfect alignment to the midpoint $p(q|s,0) = e^{-\mu(s)}{\mu}^{q}/ {q!}$. Since $M_1 \check{m}_2$ is approximately a $\chi^2$ random variable, it has variance $\mathbb{V}[M_1 \check{m}_2] = 2(M_1-1)$. Consequently, the second moment estimator (variance estimator) has variance $\mathbb{V}[\check{m}_2] = 2\left(\frac{M_1-1}{M_1^2}\right)$. We choose the hyperparameters of the Gamma prior such that the mean is matched to the maximum likelihood estimate of the rate parameter and the variance is matched to the variance of $\check{m}_2(\mbf{x})$, 
\begin{align*}
    \frac{\alpha}{\lambda} &= \check{\mu}(\mbf{x}) = \frac{1}{4}(\check{m}_2(\mbf{x}) - 1),\\
    \frac{\alpha}{\lambda^2} &= \mathbb{V}[\check{\mu}] =\frac{1}{16}\mathbb{V}[\check{m}_2] = \frac{1}{8}\bigg(\frac{M_1-1}{M_1^2}\bigg).
\end{align*}
Solving these simultaneous equations we get,
\begin{align*}
\alpha &= \frac{1}{2}(\check{m}_2(\mbf{x})-1)^2\bigg(\frac{M_1^2}{M_1 -1}\bigg),\\
\lambda &= 2(\check{m}_2(\mbf{x})-1)\bigg(\frac{M_1^2}{M_1 -1}\bigg).
\end{align*}
Since the Gamma distribution requires $\alpha,\lambda>0$, we set $\check{m}_1(\mbf{x})-1$ to a small $0<\epsilon<<1$ if $\check{m}_2(\mbf{x})<1$. We thus construct a prior on $\mu$ given by,
$$
\tilde{p}(\mu) = \text{Gamma}(\mu|\alpha,\lambda),
$$
such that the prior on the separation can be expressed as,
$$
\tilde{p}(s) = \Gamma\big((s/2\sigma)^2|\alpha,\beta) \cdot \frac{s}{2\sigma^2}.
$$

\section{Approximate Brightness Posterior}
\label{apd: Sensing Stage DI Likelihood}
Numerically computing the conditional posterior of the brightness bias parameter $\kappa$ in Eqn. \ref{eqn: Brightness Posterior} may be challenging  when integrating large numbers of photons because the direct imaging likelihood can easily drop below floating point precision. Here we use an approximation which is valid in the sub-diffraction regime to guarantee the stability of the posterior calculation. 

It is straightforward to show that, in the sub-diffraction limit $s \ll \sigma$, the direct imaging probability density for a single photon can be approximated by,
$$
p(x'|\varepsilon,s,\kappa) \approx p_0(x'-\varepsilon)\bigg[1+\frac{2\kappa s}{\sigma^2}(x'-\varepsilon)\bigg],
$$ 
where $ p_0(x) = |\psi(x)|^2 = \exp(-x^2/2\sigma^2)/\sqrt{2\pi\sigma^2}$ is the PDF of PSF. This property is extremely useful because now we may factor the direct imaging likelihood as,
\begin{equation}
p(\mbf{x}'|\varepsilon,s,\kappa) \approx \bigg(\prod_{i=1}^{N_1} p_0(x_i'-\varepsilon)\bigg)\bigg(\prod_{i=1}^{N_1}[1+\frac{2\kappa s}{\sigma^2}(x'_i-\varepsilon)]\bigg).
\end{equation}
Substituting this approximation for likelihood into Eqn. \ref{eqn: Brightness Posterior}, we find that,
$$
p(\kappa|s,\varepsilon) \approx \frac{\bigg(\prod_{i=1}^{N_1}[1+\frac{2\kappa s}{\sigma^2}(x'_i-\varepsilon)]\bigg)\tilde{p}(\kappa)}{\bigg(\int_{-\infty}^{\infty}\prod_{i=1}^{N_1}[1+\frac{2\kappa s}{\sigma^2}(x'_i-\varepsilon)] \tilde{p}(\kappa) d\kappa \bigg) },
$$
which is numerically stable for large $N_1$. With this approximation, the posterior will not suffer from a vanishingly small likelihood in the limit of many photons.

\end{document}